\documentclass[twocolumn,trackchanges]{aastex61}
\submitjournal{ApJ}

\usepackage{hyperref}
\usepackage{amsmath}
\usepackage{amssymb}
\usepackage{graphicx}

\newcommand{\Teff}{T_{\rm eff}}
\newcommand{\logg}{\log g}

\begin{document}
\title{
Chemical Abundances in the Ultra-Faint Dwarf Galaxies Grus~I and Triangulum~II:
Neutron-Capture Elements as a Defining Feature of the Faintest Dwarfs
\footnote{This paper includes data gathered with the 6.5 meter Magellan Telescopes located at Las Campanas Observatory, Chile.}}
\shorttitle{Gru~I and Tri~II}

\newcommand{\affilcarnegie}{The Observatories of the Carnegie Institution for Science, 813 Santa Barbara St., Pasadena, CA 91101, USA}
\newcommand{\affilmit}{Department of Physics and Kavli Institute for Astrophysics and Space Research, Massachusetts Institute of Technology, Cambridge, MA 02139, USA}
\newcommand{\affiluvic}{Department of Physics and Astronomy, University of Victoria, Victoria, BC V8W 3P2, Canada}

\correspondingauthor{Alexander P. Ji}
\email{aji@carnegiescience.edu}

\author{Alexander P. Ji}
\altaffiliation{Hubble Fellow}
\affiliation{\affilcarnegie}

\author{Joshua D. Simon}
\affiliation{\affilcarnegie}

\author{Anna Frebel}
\affiliation{\affilmit}

\author{Kim A. Venn}
\affiliation{\affiluvic}

\author{Terese T. Hansen}
\affiliation{\affilcarnegie}

\begin{abstract}
We present high-resolution spectroscopy of four stars in two candidate ultra-faint dwarf galaxies (UFDs), Grus~I (Gru~I) and Triangulum~II (Tri~II).
Neither object currently has a clearly determined velocity dispersion, placing them in an ambiguous region of parameter space between dwarf galaxies and globular clusters.
No significant metallicity difference is found for the two Gru~I stars, but both stars are deficient in neutron-capture elements.
We verify previous results that Tri~II displays significant spreads in metallicity and $\mbox{[$\alpha$/Fe]}$.
Neutron-capture elements are not detected in our Tri~II data, but we place upper limits at the lower envelope of Galactic halo stars, consistent with previous very low detections.
Stars with similarly low neutron-capture element abundances are common in UFDs, but rare in other environments.
This signature of low neutron-capture element abundances traces chemical enrichment in the least massive star-forming dark matter halos, and further shows that the dominant sources of neutron-capture elements in metal-poor stars are rare.
In contrast, all known globular clusters have similar ratios of neutron-capture elements to those of halo stars, suggesting that globular cluster do not form at the centers of their own dark matter halos.
The low neutron-capture element abundances may be the strongest evidence that Gru~I and Tri~II are (or once were) galaxies rather than globular clusters, and we expect future observations of these systems to robustly find non-zero velocity dispersions or signs of tidal disruption.
However, the nucleosynthetic origin of this low neutron-capture element floor remains unknown.
\end{abstract}

\section{Introduction}

Ultra-faint dwarf galaxies (UFDs) are the least luminous galaxies known.
They have only been discovered relatively recently, after the advent of deep, wide-area photometric surveys like the Sloan Digital Sky Survey, Pan-STARRS, and the Dark Energy Survey found several low surface-brightness satellites of the Milky Way \citep[e.g.,][]{Willman05a,Belokurov07,Laevens15a,Bechtol15,Koposov15a}.
Though at first it was unclear if such objects were dwarf galaxies or globular clusters \citep{Willman05a}, subsequent spectroscopic followup found most of them displayed velocity dispersions implying mass-to-light ratios $>100$ and large metallicity spreads \citep[e.g.,][]{Simon07}.
These properties contrast with globular clusters, which display no evidence for dark matter or large metallicity spreads \citep{Willman12}.

UFDs are now understood to be the natural result of galaxy formation in small dark matter halos in standard $\Lambda$CDM cosmology.
Theoretically, these galaxies begin forming at $z \sim 10$ in small $\sim 10^8~M_\odot$ dark matter halos \citep{Bromm11}.
Supernova feedback is especially effective in these small galaxies \citep[e.g.,][]{BlandHaw15},
so they form stars inefficiently for $1-2$ Gyr before their star formation is quenched by reionization \citep{Bullock00,Benson02}.
All observed properties of UFDs are also consistent with this picture.
Color-magnitude diagrams show they contain uniformly old stellar populations \citep{Brown14,Weisz14a}.
Spectroscopy shows their stars have low metallicities that extend the mass-metallicity relation all the way to $M_\star \sim 1000 M_\odot$ \citep{Kirby08,Kirby13b}.
At such tiny stellar masses, the chemical abundances of individual UFDs will not even sample a full initial mass function's worth of supernovae \citep[e.g.,][]{Koch08,Simon10,Lee13}, let alone rarer nucleosynthesis events like neutron star mergers \citep{Ji16b}.

Given the likely association between UFDs and small scale dark matter substructure, it is extremely important to distinguish between UFDs and globular clusters.
Currently, the largest telescopes can perform spectroscopy to establish velocity and metallicity dispersions from a reasonable number of stars in the closest and/or most luminous UFDs \citep[e.g.,][]{Simon07}.
However, many of the most recently discovered UFDs are very faint and/or far away. 
In such cases, only a handful of stars are accessible for followup spectroscopy, so it is difficult to clearly establish a velocity or metallicity dispersion for these galaxy candidates \citep[e.g.,][]{Koch09,Koposov15b,Kirby15,Kirby15b,Martin16,Martin16b}.
Exacerbating this concern is the presence of unresolved binary stars, which can inflate velocity dispersions and can therefore lead to premature UFD classifications \citep{McConnachie10,Ji16a,Kirby17}.
As a result, many UFD candidates still do not have clear velocity and/or metallicity dispersions \citep{Kirby15,Kirby17,Martin16,Martin16b,Walker16,Simon17}.

For some UFDs, an alternative is to examine the detailed chemical abundances of the brightest stars.
The first high-resolution spectroscopic abundances of stars in UFDs revealed that most elemental abundances in UFDs follow the average trends defined by metal-poor Milky Way halo stars, with the obvious exception of neutron-capture elements (e.g. Sr, Ba, Eu) that were extremely \emph{low} \citep{Koch08,Koch13,Frebel10b,Frebel14,Simon10}.
This view was recently revised by the discovery that some UFDs (Reticulum~II and Tucana~III) have extremely high abundances of neutron-capture elements synthesized in the $r$-process \citep{Ji16b,Roederer16b,Hansen17}.
In stark contrast, neutron-capture elements in globular clusters closely follow the abundance trends of the Milky Way halo \citep[e.g.,][]{Gratton04,Gratton12,Pritzl05}, including the globular clusters that display some internal neutron-capture abundance scatter \citep{Roederer11b}.
Extreme neutron-capture element abundances have thus been suggested to be a distinguishing factor between UFDs and globular clusters \citep{Frebel15}.

Here we study the detailed chemical abundances of the dwarf galaxy candidates Grus~I (Gru~I) and Triangulum~II (Tri~II).
Gru~I was discovered in Dark Energy Survey data by \citet{Koposov15a}.
\citet{Walker16} identified seven likely members of this galaxy, but did not resolve a metallicity or velocity dispersion.
Tri~II was discovered by \citet{Laevens15a} in Pan-STARRS.
As one of the closest but also least luminous galaxy candidates ($d_\odot = 28.4$ kpc, $M_V = -1.2$; \citealt{Carlin17}), Tri~II has already been the subject of numerous spectroscopic studies \citep{Kirby15,Kirby17,Martin16,Venn17}.
We report the first detailed chemical abundances of two stars in Gru~I and a reanalysis of two stars in Tri~II with additional data.
We describe our observations and abundance analysis in Sections~\ref{s:obs} and \ref{s:analysis}.
Section~\ref{s:abunds} details the results for individual elements.
We consider the classification of Gru~I and Tri~II in Section~\ref{s:discussion}, with an extended discussion of the origin and interpretation of neutron-capture elements in UFDs, larger dSph satellites, and globular clusters.
We conclude in Section~\ref{s:conclusion}.
\vspace{1cm}

\section{Observations and Data Reduction} \label{s:obs}

\begin{figure*}
\centering
\includegraphics[height=8cm]{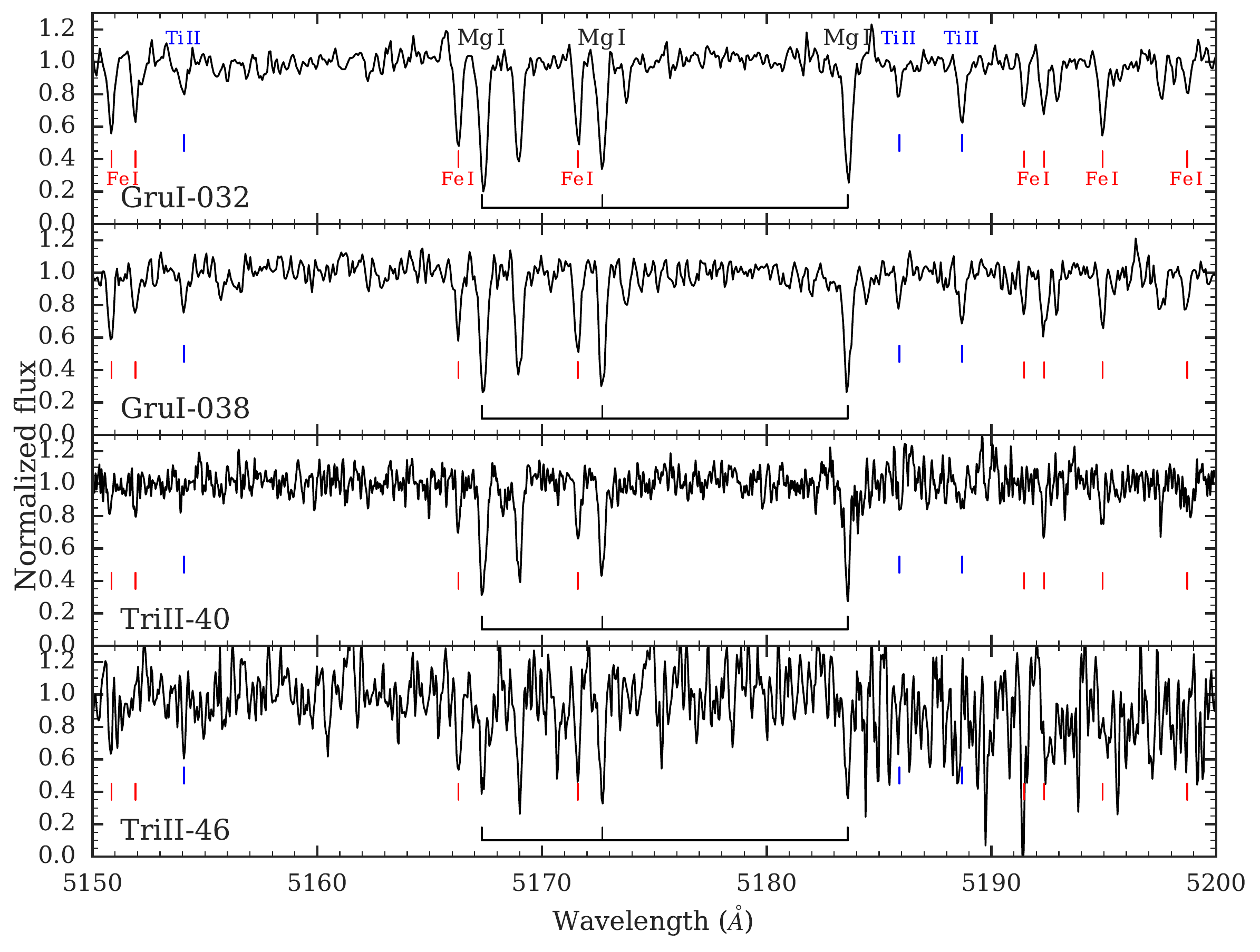}
\includegraphics[height=8cm]{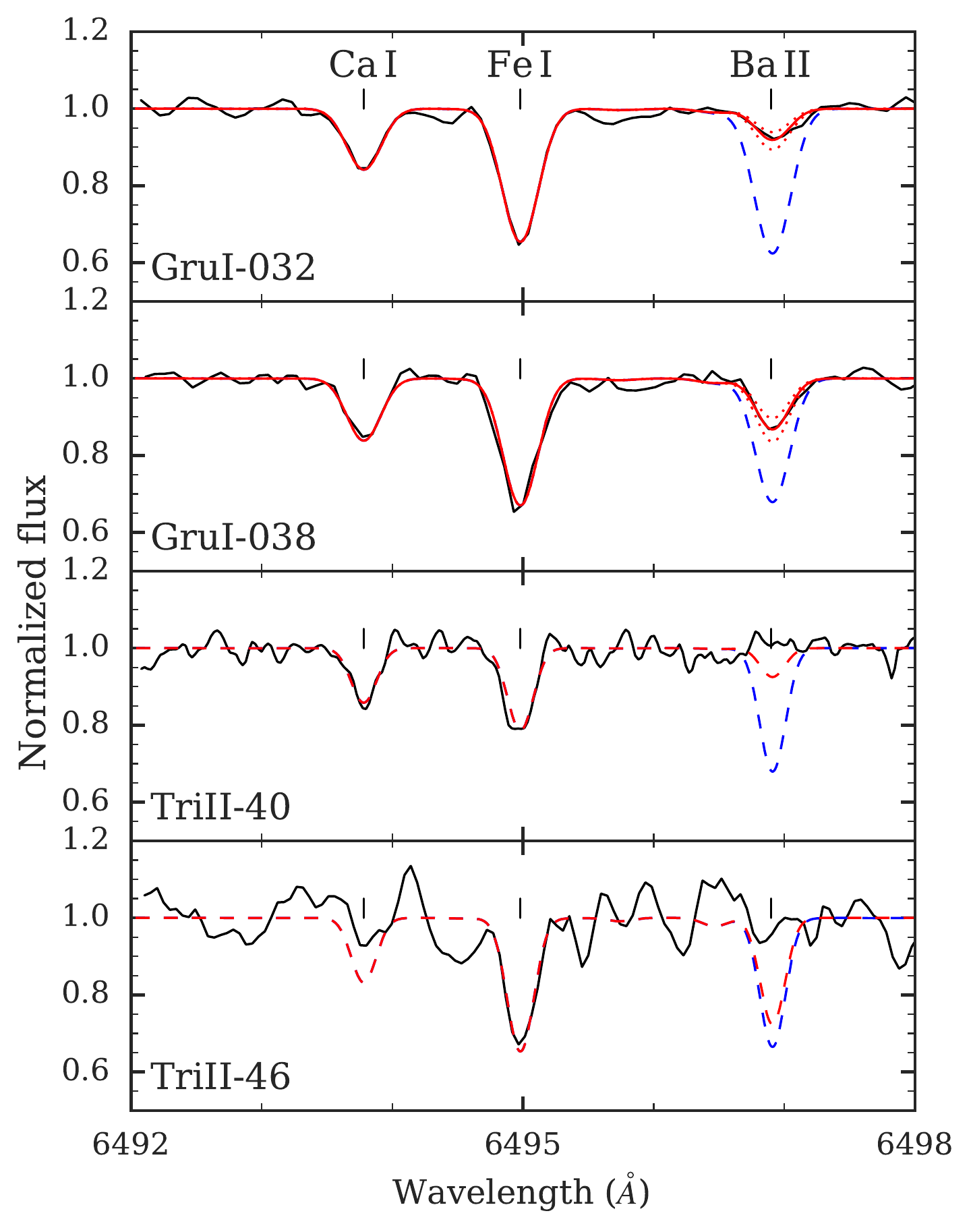}
\caption{
\emph{Left panels}: Spectra of the target stars around the Mg b triplet. Mg b, Ti II, and Fe I lines are labeled in black, blue, and red, respectively. Notice the large drop in S/N in TriII-46 at the red end due to reaching the order edge.
\emph{Right panels}: our four stars near the Ba line at 6497{\AA}. For Gru~I stars, the solid red curve indicates our best-fit synthesis, while the dotted red curves indicate $\pm 0.15$ dex. For Tri~II stars, the dashed red curves indicate upper limits. In all panels, the dashed blue line indicates $\mbox{[Ba/Fe]} = 0$ for comparison.
\label{f:spec}}
\end{figure*}

Our program stars were observed from two telescopes with two different echelle spectrographs.
Details of the observations can be found in Table~\ref{tbl:obs}.
Selected spectral regions of these four stars are shown in Figure~\ref{f:spec}.

The Gru~I stars were selected as the two brightest probable members of Gru~I from \citet{Walker16}.
We observed these stars with the Magellan Inamori Kyocera Echelle (MIKE) spectrograph \citep{Bernstein03} on the Magellan-Clay telescope in Aug 2017 with the 1\farcs0 slit, providing resolution $R \sim 28,000$ from ${\sim}3900-5000${\AA} on the blue arm and $R \sim 22,000$ from ${\sim}5000-9000${\AA} on the red arm. Individual exposures were 50-55 minutes long.
The data were reduced with CarPy \citep{Kelson03}.
Heliocentric corrections were determined with \texttt{rvcor} in IRAF\footnote{IRAF is distributed by the National Optical Astronomy Observatory, which is operated by the Association of Universities for Research in Astronomy (AURA) under a cooperative agreement with the National Science Foundation.}.

The two stars in Tri~II were observed with the Gemini Remote Access to CFHT ESPaDOnS Spectrograph (GRACES) \citep{Donati03, Chene14}\footnote{See \url{http://www.gemini.edu/sciops/instruments/visiting/graces} for more details}. These stars were selected as the brightest probable members of Tri~II from \citet{Kirby15} and \citet{Martin16}.
We combined data from two programs\footnote{GN-2015B-DD-2 (PI Venn) and GN-2016B-Q-44 (PI Ji)} that both used the 2-fiber object+sky GRACES mode providing $R \sim 40,000$ from ${\sim}5000-10,000${\AA}.
The GRACES throughput for these faint stars was worse than predicted by the integration time calculator, especially at wavelengths $<6000${\AA} where the signal-to-noise ratio (S/N) was less than half that expected.
The data were reduced with the OPERA pipeline for ESPaDOnS that was adapted for GRACES \citep{Martioli12}.
This pipeline automatically includes a heliocentric velocity correction.

\begin{deluxetable*}{lccrcrlrrrl}
\tablecolumns{11}
\tablewidth{0pt} 
\tabletypesize{\footnotesize}
\tablecaption{Observing Details\label{tbl:obs}}
\tablehead{
\colhead{Star} & \colhead{$\alpha$} & \colhead{$\delta$} & \colhead{$V$} & \colhead{Observation Date} & \colhead{$t_{\rm exp}$} & \colhead{$v_{\rm hel}$} & \colhead{S/N} & \colhead{S/N} &\colhead{S/N} & \colhead{Instrument}\\
\colhead{} & \colhead{(J2000)} & \colhead{(J2000)} & \colhead{(mag)} & \colhead{} & \colhead{(min)} & \colhead{(km s$^{-1}$)} & \colhead{(4500\AA)} & \colhead{(5300\AA)} & \colhead{(6500\AA)} & \colhead{}
}
\startdata
GruI-032 & 22 56 58.1 & $-$50 13 57.9 & 18.1 & 2017 Aug 16,25 & 165 & $-139.8 \pm 0.7$ & 22 & 25 & 60 & MIKE 1\farcs0 slit\\
GruI-038 & 22 56 29.9 & $-$50 04 33.3 & 18.7 & 2017 Aug 15,16,25 & 430 & $-143.9 \pm 0.4$ & 20 & 22 & 55 & MIKE 1\farcs0 slit\\
TriII-40 & 02 13 16.5 & $+$36 10 45.9 & 17.3 & 2015 Dec 15 & 60 & $-381.5 \pm 1.3$ &  5 & 15 & 35 & GRACES  2-fiber\\
 & & & & 2016 Sep 8 &  80 & $-381.5$ & & & & GRACES 2-fiber\\
TriII-46 & 02 13 21.5 & $+$36 09 57.6 & 18.8 & 2015 Dec 16,17 & 160 & $-396.5 \pm 3.2$ &  1 &  7 & 17 & GRACES 2-fiber\\
 & & & & 2016 Sep 7 & 120 & $-381.5 \pm 5.0$ & & & & GRACES 2-fiber\\
\enddata
\tablecomments{S/N values are per pixel. S/N values for Tri~II stars were determined after coadding. Velocity precision is computed with coadded spectra except for TriII-46, where each visit is measured separately because of the binary orbital motion.}
\end{deluxetable*}

We used IRAF and SMH \citep{Casey14} to coadd, normalize, stitch orders, and Doppler correct the reduced spectra.
We estimated the S/N per pixel on coadded spectra by running a median absolute deviation filter across the normalized spectra in a ${\approx}$5{\AA} window.
The signal-to-noise at the order center closest to rest wavelengths of 4500{\AA}, 5300{\AA}, and 6500{\AA} is given in Table~\ref{tbl:obs}.
Radial velocities were determined by cross correlating the Mg b triplet against a MIKE spectrum of HD122563.
\citet{Venn17} found one of the stars in Tri~II (TriII-46) to be a binary, so we Doppler shifted spectra from each visit to rest frame before coadding.
The implications of this binary star were previously considered in \citet{Venn17} and \citet{Kirby17}. Our added velocity measurement does not affect their conclusions.

Other than TriII-46, the velocities are consistent with constant heliocentric velocity in our data and with previous velocity measurements \citep{Kirby15,Kirby17,Martin16,Walker16,Venn17}.
Velocity precision was estimated using the coadded spectra by cross-correlating all orders from $5000-6500${\AA} for MIKE and $4500-6500${\AA} for GRACES against HD122563.
We excluded orders where the velocity was not within 10~km~s$^{-1}$ of the Mg b velocity, then took the standard deviation of the remaining order velocities. 
This value was added in quadrature to the combined statistical velocity uncertainty to obtain the velocity uncertainties listed in Table~\ref{tbl:obs}.
The most discrepant velocity other than TriII-46 is for GruI-032, which is ${\approx}1$~km~s$^{-1}$ away from the measurement in \citet{Walker16} ($-138.4 \pm 0.4$~km~s$^{-1}$), but not large enough that we would consider this a clear binary candidate.
Note that the two Gru~I stars differ by ${\approx}4$ km s$^{-1}$, which could be consistent with a significant velocity dispersion.

\section{Abundance Analysis}\label{s:analysis}

We analyzed all four stars using the 2011 version of the 1D LTE radiative transfer code MOOG \citep{Sneden73, Sobeck11} with the \citet{Castelli04} (ATLAS) model atmospheres.
We measured equivalent widths and ran MOOG with SMH \citep{Casey14}.
The abundance of most elements was determined from equivalent widths.
We used spectral synthesis to account for blends, molecules, and hyperfine structure for the species CH, Sc, Mn, Sr, Ba, and Eu.
Atomic data references can be found in table~3 of \citet{Roederer10}.
Measurements and uncertainties of individual features are in Table~\ref{tbl:eqw}.
Stellar parameters and uncertainties for this work and previous measurements are in Table~\ref{tbl:sp}.
Final abundances and uncertainties are in Table~\ref{tbl:abunds}.
Detailed abundance uncertainties due to stellar parameter variations are in Table~\ref{tbl:spabunderr}.

\begin{deluxetable*}{llrrrrrrrrrr}
\tablecolumns{20}
\tablewidth{0pt}
\tabletypesize{\footnotesize}
\tabletypesize{\tiny}
\tablecaption{\label{tbl:eqw}}
\tablehead{\colhead{El.} & \colhead{$\lambda$} & \colhead{$\chi$} & \colhead{$\log gf$} & \colhead{EW} & \colhead{$\sigma(EW)$} & \colhead{$\log \epsilon$} & \colhead{$\sigma(\log\epsilon)$} & \colhead{EW} & \colhead{$\sigma(EW)$} & \colhead{$\log \epsilon$} & \colhead{$\sigma(\log\epsilon)$}\\
\colhead{} & \colhead{(\AA)} & \colhead{(eV)} & \colhead{(dex)} & \multicolumn{4}{c}{GruI-032} & \multicolumn{4}{c}{TriII-40}}
\startdata
CH    &  4313.00 &\nodata&  \nodata &     syn & \nodata &    5.45 &    0.20 & \nodata & \nodata & \nodata & \nodata \\
CH    &  4323.00 &\nodata&  \nodata &     syn & \nodata &    5.25 &    0.30 & \nodata & \nodata & \nodata & \nodata \\
Na I  &  5889.95 &  0.00 & $  0.11$ &   210.7 &    29.2 &    3.59 &    0.37 &   104.9 &    16.7 &    2.49 &    0.25 \\
Na I  &  5895.92 &  0.00 & $ -0.19$ &   175.4 &    25.5 &    3.43 &    0.36 &    83.9 &    14.6 &    2.51 &    0.20 \\
\enddata
\tablecomments{The full version of this table is available online. A portion is shown here for form and content.}
\end{deluxetable*}

\subsection{Standard analysis for brighter stars}
For three of our stars (GruI-032, GruI-038, and TriII-40), our spectra are of sufficient quality for a standard equivalent width analysis.
We first fit Gaussian profiles to the line list in \citet{Roederer10}.
We applied the formula from \citet{Battaglia08} to determine equivalent width uncertainties.
The S/N per pixel was calculated with median absolute deviation in a running 5{\AA} window.
Varying the window size affected the S/N estimates by only 2-3\%, but we conservatively add an additional 10\% uncertainty to each equivalent width.
Using this estimate, we rejected most lines with equivalent width uncertainties larger than 30\%.
The exceptions were lines of Al, Si, Cr, Co, and Zn that otherwise would have had all lines of that element rejected; and some clean lines near regions of large true variation (e.g., near CH bands) where the S/N was clearly underestimated.
We propagate these to a $1\sigma$ abundance uncertainty for each line (Table~\ref{tbl:eqw}).
Synthesis uncertainties are calculated by varying abundances until the entire synthesized profile encompasses the spectrum noise around the feature, corresponding to $1\sigma$ uncertainties.

We derived the effective temperature, surface gravity, and microturbulence ($\Teff$, $\logg$, $\nu_t$) with excitation, ionization, and line strength balance of Fe lines. We then applied the $\Teff$ correction from \citet{Frebel13} and redetermined $\logg$ and $\nu_t$.
Statistical uncertainties for $\Teff$ and $\nu_t$ correspond to the $1\sigma$ error on the fitted slopes of abundance with respect to excitation potential and reduced equivalent width, respectively.
The statistical uncertainty for $\logg$ was derived by varying the parameter to match the combined standard error of the Fe\,I and Fe\,II abundances.
We then further adopt systematic uncertainties of 150 K for $\Teff$ from scatter in the \citet{Frebel13} calibration; and 0.3 dex for $\logg$, and 0.3 km s$^{-1}$ for $\nu_t$ to reflect this systematic temperature uncertainty.
We use the standard deviation of Fe\,I lines as the statistical uncertainty in the stellar atmosphere's model metallicity.
We add the statistical and systematic uncertainties in quadrature to obtain the stellar parameter uncertainties in Table~\ref{tbl:sp}.
These three stars are all $\alpha$-enhanced, so we used the $\mbox{[$\alpha$/Fe]} = +0.4$ \citet{Castelli04} model atmospheres.

\subsection{Analysis of TriII-46}
The data for star TriII-46 has very low signal-to-noise (Table~\ref{tbl:obs}) and thus requires special care.
We rebin the spectra by a factor of 2 to improve the signal-to-noise.
This allowed us to measure equivalent widths for lines at the center of echelle orders with wavelengths $>5000${\AA}.
After keeping only lines with equivalent width uncertainty less than 30\%, we have 18 Fe\,I lines and only one Fe\,II line.

For this small number of lines, spectroscopic determination of stellar parameters is subject to many degeneracies based on line selection.
Still, we examine here what parameters would be derived with the information from Fe lines.
If we apply the same procedure as for the other three stars (i.e., excitation, ionization, and line strength balance with the \citealt{Frebel13} correction but using only these 19 lines), we obtain $\Teff=5260$\,K, $\logg=2.1$\,dex, $\nu_t=2.60$\,km/s, and $\mbox{[Fe/H]} = -2.01$.
However, the ionization equilibrium is set by a single Fe\,II line with equivalent width $164 \pm 50$ m{\AA}, so this is extremely unreliable.
Ignoring the Fe\,II line and using a $\mbox{[Fe/H]}=-2$ Yonsei-Yale isochrone to set $\logg$ as a function of $\Teff$ \citep{Kim02}, we obtain $\Teff=5260$\,K, $\logg=2.7$\,dex, $\nu_t=2.50$\,km/s, and $\mbox{[Fe/H]} = -2.01$.
The statistical errors are large: 240K, 0.6 dex, 0.5 km/s, and 0.3 dex respectively.

We summarize this and other derived stellar parameters for this star in Table~\ref{tbl:sp}.
For comparison, \citet{Venn17} derived $\Teff=5050$\,K, $\logg=2.6$\,dex, and $\nu_t=2.5$\,km/s for TriII-46 using photometry, distance, and a modified scaling relation for $\nu_t$.
An updated distance modulus \citep{Carlin17} would slightly increase $\logg$ to 2.7\,dex.
\citet{Kirby17} derived $\Teff=5282$\,K, $\logg=2.74$\,dex, and $\nu_t=1.5$\,km/s using photometry and distance to set $\logg$ and $\nu_t$ but allowing $\Teff$ to vary to fit their spectrum.
Our stellar parameters are somewhat in between their values, preferring the higher temperature from \citet{Kirby17} but with the higher microturbulence from \citet{Venn17}.
Our data for this star are insufficient to make any further refinements, so we decided to adopt intermediate values with large uncertainties that encompass other stellar parameter determinations: $\Teff=5150 \pm 200$\,K, $\logg=2.7 \pm 0.5$\,dex, and $\nu_t=2.0 \pm 0.5$\,km/s.
Regardless of the stellar parameters, this star is not $\alpha$-enhanced so we use the \citet{Castelli04} model atmospheres with $\mbox{[$\alpha$/Fe]} = 0$.
We propagate these uncertainties through to the final abundance uncertainties.

\subsection{Final Abundances and Uncertainties}
Table~\ref{tbl:abunds} contains the final abundance results for our stars.
For each element, $N$ is the number of lines measured.
$\log \epsilon(X)$ is the average abundance of those lines weighted by the abundance uncertainty.
Letting $\log\epsilon_i$ and $\sigma_i$ be the abundance and uncertainty of line~$i$, we define $w_i = 1/\sigma_i^2$ and $\log\epsilon(X) = \sum_i (w_i \log\epsilon_i) / \sum_i w_i$.
$\sigma$ is the standard deviation of those lines.
$\sigma_{\rm w}$ is the standard error from propagating individual line uncertainties, i.e., $1/\sigma_{\rm w}^2 = \sum_i w_i$ \citep{McWilliam95}.
$\mbox{[X/H]}$ is the abundance relative to solar abundances from \citet{Asplund09}.
$\mbox{[X/Fe]}$ is calculated using either [Fe\,I/H] or [Fe\,II/H], depending on whether X is neutral or ionized; except for TriII-46, where all [X/Fe] are calculated relative to [Fe\,I/H] because of an unreliable Fe\,II abundance.
$\sigma_{\rm [X/H]}$ is the quadrature sum of $\sigma/\sqrt{N}$, $\sigma_{\rm w}$, and abundance uncertainties due to $1\sigma$ stellar parameter variations.
Detailed abundance variations from changing each stellar parameter are given in Table~\ref{tbl:spabunderr}.
$\sigma_{\rm [X/Fe]}$ is similar to $\sigma_{\rm [X/H]}$, but when calculating the stellar parameter uncertainties we include variations in Fe.
We use the difference in Fe\,I abundance for neutral species and the difference in Fe\,II abundance for ionized species to calculate this error.
The [X/Fe] error is usually smaller than the [X/H] error, since abundance differences from changing $\Teff$ and $\logg$ usually (but not always) affect X and Fe in the same direction when using the same ionization state.
Since most of our elements have very few lines, we adopt the standard deviation of the Fe\,I lines as the minimum $\sigma$ when calculating $\sigma_{\rm [X/H]}$ and $\sigma_{\rm [X/Fe]}$.

Upper limits were derived by spectrum synthesis.
Using several features of each element (Table~\ref{tbl:eqw}), we found the best-fit synthesis to the observed spectrum to determine a reference $\chi^2$ and smoothing for the synthetic spectrum.
The minimum smoothing was calculated using $\rm{FWHM} = \lambda/R$ where $\lambda$ is the line wavelength.
Holding the continuum, smoothing, and radial velocity fixed, we increased the abundance until $\Delta\chi^2 = 25$. This is formally a $5\sigma$ upper limit, though it does not include uncertain continuum placement.

\begin{deluxetable}{lcrrllllll}
\tablecolumns{12}
%\tablewidth{0pt} 
\tablewidth{\linewidth}
\tabletypesize{\footnotesize}
%\tabletypesize{\tiny}
\tablecaption{Stellar Parameters\label{tbl:sp}}
\tablehead{
\colhead{Star} & \colhead{Ref} & \colhead{$\Teff$} & \colhead{$\sigma$} & \colhead{$\logg$} & \colhead{$\sigma$} & \colhead{$\nu_t$}\tablenotemark{a} & \colhead{$\sigma$} & \colhead{[Fe/H]} & \colhead{$\sigma$}}
\startdata
GruI-032 & TW & 4495 & 155 & 0.85 & 0.37 & 2.60 & 0.32 & $-$2.57 & 0.19 \\
GruI-032 &W16\tablenotemark{b} & 4270 & 69 & 0.72 & 0.22 & 2.0\   & \nodata & $-$2.69 & 0.10 \\
\hline
GruI-038 & TW & 4660 & 158 & 1.45 & 0.39 & 2.40 & 0.32 & $-$2.50 & 0.24 \\
GruI-038 &W16\tablenotemark{b} & 4532 & 100 & 0.87 & 0.31 & 2.0   & \nodata & $-$2.42 & 0.15 \\
\hline
TriII-40 & TW & 4720 & 175 & 1.35 & 0.42 & 2.48 & 0.34 & $-$2.95 & 0.21 \\
TriII-40 &V17 & 4800 & 50 & 1.80 & 0.06 & 2.7  & 0.2 & $-$2.87 & 0.19 \\
TriII-40 &K17\tablenotemark{c} & 4816 & \nodata & 1.64 & \nodata & 2.51 & \nodata & $-$2.92 & 0.21 \\
TriII-40 &K17\tablenotemark{d} & 4917 & \nodata & 1.89 & \nodata & 1.70 & \nodata & $-$2.78 & 0.11 \\
\hline
TriII-46 & TW & 5150 & 200 & 2.7 & 0.5 & 2.00 & 0.5 & $-$1.96 & 0.28 \\
TriII-46 &V17 & 5050 & 50 & 2.60 & 0.06 & 2.5  & \nodata & $-$2.5  & 0.2 \\
TriII-46 &K17\tablenotemark{d} & 5282 & \nodata & 2.74 & \nodata & 1.50 & \nodata & $-$1.91 & 0.11 \\
TriII-46 &Spec\tablenotemark{e} & 5260 & 240 & 2.7 & 0.6 & 2.5 & 0.5 & $-$2.01 & 0.26 \\
\enddata
\tablerefs{TW = this work; W16 = \citealt{Walker16}; V17 = \citealt{Venn17}; K17 = \citealt{Kirby17}}
\tablenotetext{a}{$\nu_t$ for W16 is always 2 km/s \citep{Lee08a}. $\nu_t$ for DEIMOS data in K17 according to the equation $\nu_t = 2.13 - 0.23 \logg$ \citep{Kirby09}}
\tablenotetext{b}{[Fe/H] for W16 stars have a 0.32 dex offset removed; see text}
\tablenotetext{c}{HIRES data}
\tablenotetext{d}{DEIMOS data}
\tablenotetext{e}{Spectroscopic balances in this work using isochrones to determine $\logg$}
\end{deluxetable}

\subsection{Comparison to literature measurements}\label{s:litcomp}
For the two Gru~I stars, \citet{Walker16} determined stellar parameters and metallicities from high-resolution M2FS spectra near the Mg b triplet using a large synthesized grid. The grid fixes $\nu_t = 2.0$ \citep{Lee08a}. \citet{Walker16} increased all their [Fe/H] measurements by $0.32$ dex, which is the offset they obtained from fitting twilight spectra of the Sun. It is not clear that the same offset should be applied for both dwarf stars (like the Sun) and giants. If we remove the offset, our stellar parameters and metallicities are in good agreement (also see \citealt{Ji16d}).

\citet{Venn17} analyzed both stars in Tri~II, and we have combined their previous GRACES data with additional observations\footnote{\citet{Venn17} labeled the stars as Star~40 and Star~46 instead of TriII-40 and TriII-46. We have retained the number but changed the label to TriII for clarity.}.
For TriII-40, we find good agreement for all stellar parameters except $\logg$. This is because we determined our $\logg$ spectroscopically, while \citet{Venn17} did so photometrically using the distance to Tri~II.
Adjusting for the different $\logg$, our abundances for this star agree within $1\sigma$.
For TriII-46, \citet{Venn17} fixed stellar parameters with photometry and used spectral synthesis to measure all abundances.
We measured $\mbox{[Fe/H]} = -2.01 \pm 0.37$, while \citet{Venn17} obtained $\mbox{[Fe/H]} = -2.5 \pm 0.2$.
Our large abundance uncertainty means these are only $1.2 \sigma$ discrepant, but we might expect better agreement given that so much of the data overlaps.
Detailed investigation of the discrepancy shows that 0.3 dex of the difference is due to differences in stellar parameters (mostly $\Teff$ and $\nu_t$).
The remaining 0.2 dex is attributable to systematic differences in continuum placement that are individually within $1\sigma$ uncertainties.
Finally, we note that the stellar parameter uncertainties in \citet{Venn17} reflect statistical photometric errors, but could be larger due to systematic uncertainties in photometric calibrations, filter conversions, and reddening maps.

\citet{Kirby17} determined abundances of TriII-40 with a high-resolution, high signal-to-noise Keck/HIRES spectrum.
Our abundances agree within 0.15 dex, except for Cr which is still within $1\sigma$.
\citet{Kirby17} also analyzed the Mg, Ca, Ti, and Fe abundance of TriII-46 by matching a synthetic grid to an $R \sim 7000$ Keck/DEIMOS spectrum.
They measured $\mbox{[Mg/Fe]}=+0.21 \pm 0.28$, $\mbox{[Ca/Fe]}=-0.39 \pm 0.15$, and $\mbox{[Ti/Fe]}=-0.79 \pm 0.76$.
There are some significant discrepancies, especially for Mg.
One possible reason for these differences is that we used stronger blue lines with lower excitation potentials for Mg and Ti, while the synthetic grid is driven by combining multiple higher excitation potential lines that we could not individually measure in our spectrum. This explanation is supported by the fact that our Ca abundances agree better because they are derived from similar spectral features.

\section{Abundance Results}\label{s:abunds}

In Gru~I we measured the abundances of C, Na, Mg, Al, Si, K, Ca, Sc, Ti, Cr, Mn, Fe, Co, Ni, Sr, and Ba.
In Tri~II we were only able to measure Mg, K, Ca, Ti, Cr, Fe, Co, Ni, and Ba due to a combination of lower S/N and the fact that the strongest features for other elements are found $\lambda<5000${\AA}.

\begin{figure*}
\centering
\includegraphics[width=18cm]{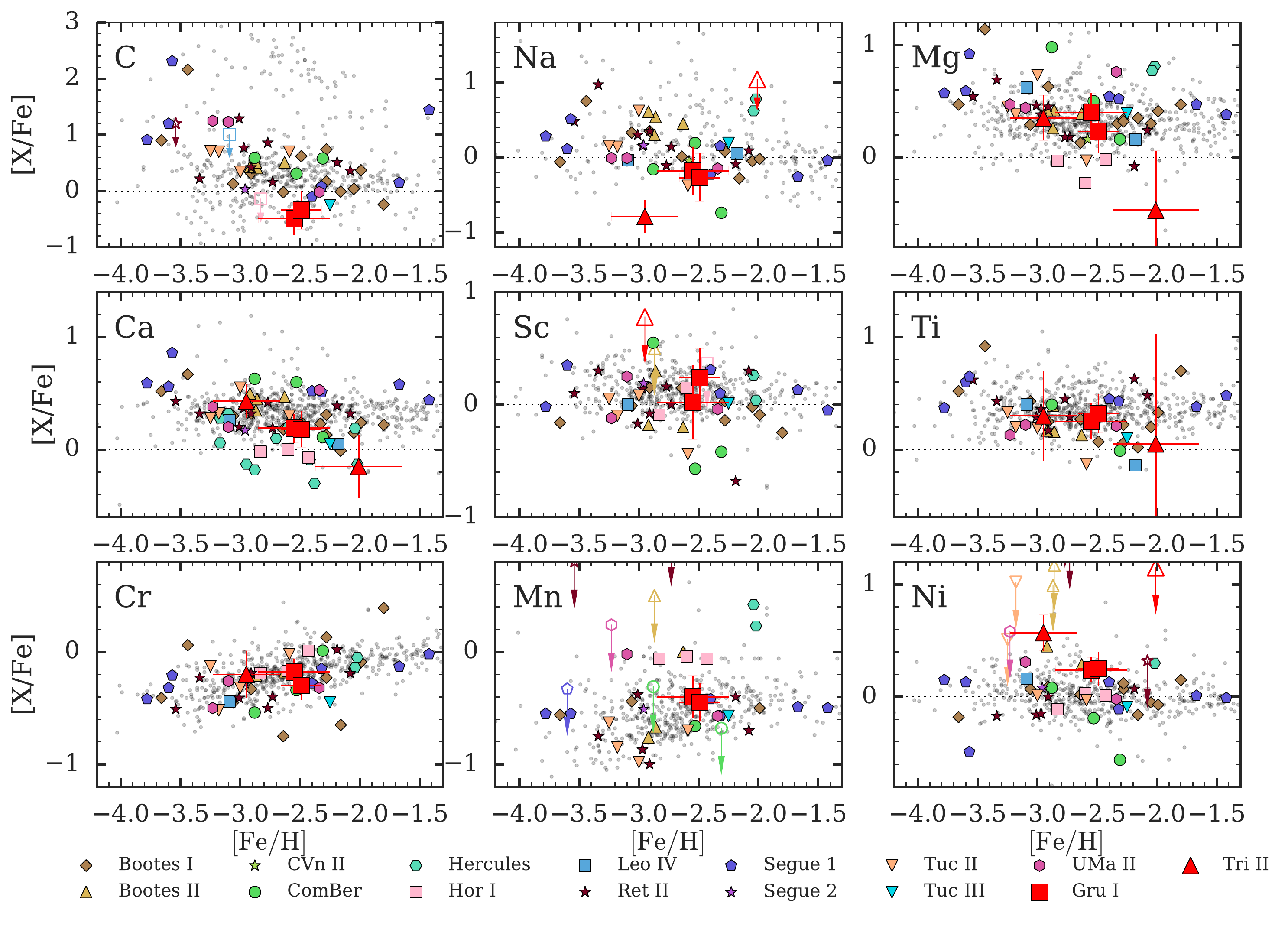}
\caption{Abundance of light elements in Gru~I (red squares) and Tri~II (red triangles) compared to halo stars (gray points) and other UFDs (colored points).
Upper limits are indicated as open symbols with arrows.
The element $X$ is indicated in the top-left corner of each panel.
Tri~II stars are not plotted for C. Limits on Tri~II abundances are above the top axis for Sc and Mn.
Essentially all [X/Fe] ratios in these two galaxies follow trends defined by the Milky Way halo stars and other UFDs. The notable exceptions are the Na and Ni in TriII-40, and the low Mg and Ca in TriII-46.
%(NOTE: I cut Al, Si, Co. These are not useful elements given the errors.)
\label{f:grid1}}
\end{figure*}

Figures~\ref{f:grid1}, \ref{f:kmg}, and \ref{f:grid2} show the abundances of our four stars compared to other UFDs and a literature sample of halo stars (\citealt{Frebel10}, and \citealt{Roederer14c} for K).
The UFDs are Bootes~I \citep{Feltzing09,Norris10a,Gilmore13,Ishigaki14,Frebel16}, Bootes~II \citep{Ji16b}, CVn~II \citep{Francois16}, Coma Berenices \citep{Frebel10b}, Hercules \citep{Koch08, Koch13}, Hor~I \citep{Nagasawa18}, Leo~IV \citep{Simon10,Francois16}, Reticulum~II \citep{Ji16c,Roederer16b}, Segue~1 \citep{Frebel14}, Segue~2 \citep{Roederer14a}, Tuc~II \citep{Ji16d,Chiti18}, Tuc~III \citep{Hansen17}, and UMa~II \citep{Frebel10b}.

Overall, the two Gru~I stars have the same [Fe/H] to within our abundance uncertainties, and all [X/Fe] ratios are very similar except for Ba.
The metallicities of the Tri~II stars differ by more than $2\sigma$ and display different abundance ratios.
We now discuss each element in more detail.

\startlongtable
\begin{deluxetable}{lrrrr|rr|rr}
\tablecolumns{9}
\tablewidth{0pt}
\tabletypesize{\footnotesize}
\tabletypesize{\tiny}
\tablecaption{Abundances\label{tbl:abunds}}
\tablehead{\colhead{Species} & \colhead{$N$} & \colhead{$\log \epsilon(X)$} & \colhead{$\sigma$} & \colhead{$\sigma_{\rm w}$} & \colhead{$\mbox{[X/H]}$} & \colhead{$\sigma_{\rm [X/H]}$} & \colhead{$\mbox{[X/Fe]}$} & \colhead{$\sigma_{\mbox{[X/Fe]}}$}}
\startdata
\cutinhead{ GruI-032 }
CH   &   2 &  5.39 & 0.14 & 0.17 & $-$3.04 & 0.40 & $-$0.49 & 0.29\\
Na I  &   2 &  3.51 & 0.11 & 0.26 & $-$2.73 & 0.50 & $-$0.18 & 0.32\\
Mg I  &   5 &  5.45 & 0.12 & 0.13 & $-$2.15 & 0.35 &  0.40 & 0.17\\
Al I  &   1 &  2.97 &\nodata&1.18 & $-$3.48 & 1.27 & $-$0.93 & 1.21\\
Si I  &   1 &  5.38 &\nodata&0.66 & $-$2.13 & 0.75 &  0.42 & 0.70\\
K I   &   2 &  3.09 & 0.03 & 0.13 & $-$1.94 & 0.31 &  0.61 & 0.20\\
Ca I  &  13 &  3.98 & 0.19 & 0.05 & $-$2.36 & 0.20 &  0.20 & 0.13\\
Sc II &   5 &  0.62 & 0.31 & 0.21 & $-$2.53 & 0.32 & $-$0.01 & 0.33\\
Ti I  &  11 &  2.53 & 0.22 & 0.07 & $-$2.42 & 0.38 &  0.13 & 0.15\\
Ti II &  26 &  2.65 & 0.19 & 0.05 & $-$2.30 & 0.19 &  0.21 & 0.16\\
Cr I  &   9 &  2.91 & 0.17 & 0.07 & $-$2.73 & 0.33 & $-$0.18 & 0.12\\
Mn I  &   7 &  2.48 & 0.30 & 0.12 & $-$2.95 & 0.38 & $-$0.40 & 0.19\\
Fe I  & 112 &  4.95 & 0.19 & 0.02 & $-$2.55 & 0.30 &  0.00 & \nodata\\
Fe II &  10 &  4.98 & 0.25 & 0.09 & $-$2.52 & 0.23 &  0.00 & \nodata\\
Co I  &   3 &  2.76 & 0.11 & 0.40 & $-$2.23 & 0.61 &  0.32 & 0.45\\
Ni I  &   7 &  3.91 & 0.18 & 0.06 & $-$2.31 & 0.29 &  0.24 & 0.11\\
Zn I  &   1 &  1.95 &\nodata&0.18 & $-$2.61 & 0.28 & $-$0.06 & 0.40\\
Sr II &   2 & $-$1.65 & 0.28 & 0.35 & $-$4.52 & 0.45 & $-$2.00 & 0.45\\
Ba II &   4 & $-$1.92 & 0.33 & 0.07 & $-$4.10 & 0.25 & $-$1.58 & 0.30\\
Eu II &   1 &$<-1.68$&\nodata&\nodata&$<-2.20$&\nodata&$<+0.32$&\nodata\\
\cutinhead{GruI-038}
CH   &   2 &  5.60 & 0.03 & 0.19 & $-$2.83 & 0.45 & $-$0.34 & 0.34\\
Na I  &   2 &  3.48 & 0.06 & 0.26 & $-$2.76 & 0.48 & $-$0.27 & 0.32\\
Mg I  &   5 &  5.34 & 0.19 & 0.12 & $-$2.26 & 0.34 &  0.23 & 0.19\\
Al I  &   1 &  3.08 &\nodata&1.42 & $-$3.37 & 1.49 & $-$0.88 & 1.44\\
Si I  &   1 &  5.56 &\nodata&0.58 & $-$1.95 & 0.69 &  0.54 & 0.63\\
K I   &   2 &  3.24 & 0.06 & 0.14 & $-$1.79 & 0.34 &  0.71 & 0.23\\
Ca I  &  11 &  4.03 & 0.26 & 0.06 & $-$2.31 & 0.21 &  0.19 & 0.16\\
Sc II &  10 &  0.90 & 0.16 & 0.10 & $-$2.25 & 0.25 &  0.21 & 0.26\\
Ti I  &   8 &  2.47 & 0.20 & 0.09 & $-$2.48 & 0.33 &  0.02 & 0.14\\
Ti II &  37 &  2.78 & 0.23 & 0.05 & $-$2.17 & 0.22 &  0.29 & 0.17\\
Cr I  &   8 &  2.85 & 0.16 & 0.09 & $-$2.79 & 0.35 & $-$0.30 & 0.13\\
Mn I  &   7 &  2.49 & 0.26 & 0.10 & $-$2.94 & 0.37 & $-$0.45 & 0.17\\
Fe I  & 107 &  5.01 & 0.24 & 0.02 & $-$2.49 & 0.31 &  0.00 & \nodata\\
Fe II &   7 &  5.04 & 0.26 & 0.11 & $-$2.46 & 0.25 &  0.00 & \nodata\\
Co I  &   1 &  2.86 &\nodata&0.74 & $-$2.13 & 0.90 &  0.36 & 0.79\\
Ni I  &   5 &  3.98 & 0.10 & 0.07 & $-$2.24 & 0.28 &  0.25 & 0.15\\
Sr II &   2 & $-$1.65 & 0.14 & 0.35 & $-$4.52 & 0.44 & $-$2.06 & 0.44\\
Ba II &   4 & $-$1.23 & 0.14 & 0.10 & $-$3.41 & 0.25 & $-$0.94 & 0.25\\
Eu II &   1 &$<-1.20$&\nodata&\nodata&$<-1.72$&\nodata&$<+0.74$&\nodata\\
\cutinhead{TriII-40}
Na I  &   2 &  2.50 & 0.01 & 0.16 & $-$3.74 & 0.33 & $-$0.79 & 0.22\\
Mg I  &   3 &  5.00 & 0.10 & 0.13 & $-$2.60 & 0.32 &  0.35 & 0.20\\
K I   &   1 &  2.89 &\nodata&0.30 & $-$2.14 & 0.36 &  0.81 & 0.31\\
Ca I  &   8 &  3.82 & 0.24 & 0.05 & $-$2.52 & 0.20 &  0.43 & 0.15\\
Sc II &   1 &$<0.98$&\nodata&\nodata&$<-2.17$&\nodata&$<+0.65$&\nodata\\
Ti I  &   3 &  2.31 & 0.08 & 0.13 & $-$2.64 & 0.32 &  0.31 & 0.19\\
Ti II &   2 &  2.30 & 0.44 & 0.18 & $-$2.65 & 0.40 &  0.17 & 0.40\\
Cr I  &   3 &  2.49 & 0.27 & 0.13 & $-$3.15 & 0.35 & $-$0.20 & 0.21\\
Mn I  &   1 &$<2.97$&\nodata&\nodata&$<-2.46$&\nodata&$<+0.36$&\nodata\\
Fe I  &  60 &  4.55 & 0.21 & 0.02 & $-$2.95 & 0.28 &  0.00 & \nodata\\
Fe II &   5 &  4.68 & 0.33 & 0.09 & $-$2.82 & 0.23 &  0.00 & \nodata\\
Ni I  &   3 &  3.84 & 0.19 & 0.09 & $-$2.38 & 0.30 &  0.57 & 0.16\\
Ba II &   1 &$<-1.89$&\nodata&\nodata&$<-4.07$&\nodata&$<-1.25$&\nodata\\
Eu II &   1 &$<-0.89$&\nodata&\nodata&$<-1.41$&\nodata&$<+1.41$&\nodata\\
\cutinhead{TriII-46}
Na I  &   1 &$<5.27$&\nodata&\nodata&$<-0.97$&\nodata&$<1.04$&\nodata\\
Mg I  &   2 &  5.12 & 0.02 & 0.42 & $-$2.48 & 0.61 & $-$0.47 & 0.53\\
K I   &   1 &$<3.79$&\nodata&\nodata&$<-1.24$&\nodata&$<0.77$&\nodata\\
Ca I  &   3 &  4.18 & 0.30 & 0.15 & $-$2.16 & 0.33 & $-$0.15 & 0.28\\
Sc II &   1 &$<2.65$&\nodata&\nodata&$<-0.50$&\nodata&$<1.51$&\nodata\\
Ti II &   1 &  2.99 &\nodata&0.38 & $-$1.96 & 0.53 & 0.05 & 0.98\\
Fe I  &  18 &  5.49 & 0.29 & 0.09 & $-$2.01 & 0.36 &  0.00 & \nodata\\
Fe II &   1 &  5.98 &\nodata&0.82 & $-$1.52 & 0.92 &  0.49 & \nodata\\
Ni I  &   1 &$<5.36$&\nodata&\nodata&$<-0.86$&\nodata&$<1.15$&\nodata\\
Ba II &   1 &$<-0.06$&\nodata&\nodata&$<-2.24$&\nodata&$<-0.23$&\nodata\\
Eu II &   1 &$<0.50$&\nodata&\nodata&$<-0.02$&\nodata&$<1.99$&\nodata\\
\enddata
\end{deluxetable}

\subsection{Carbon}
Spectral synthesis of the $G$-band features at 4313{\AA} and 4323{\AA} was used to measure carbon in the Gru~I stars (using a list from B. Plez 2007, private communication).
The oxygen abundance can affect molecular equilibrium, but since oxygen cannot be measured in these stars we assume $\mbox{[O/Fe]}=0.4$.
Since they are red giant branch stars, some C has been converted to N.
The corrections from \citet{Placco14} were applied to estimate the natal abundance, which are $\mbox{[C/Fe]} = +0.21$ and $+0.57$ for GruI-032 and GruI-038, respectively.
Varying $\logg$ by the uncertainty in Table~\ref{tbl:sp} causes the correction to change by $\pm 0.1$ dex.
Both stars are carbon-normal ($\mbox{[C/Fe]} < 0.7$) even after this carbon correction. Note that the uncorrected carbon abundances are used in Figure~\ref{f:grid1} and Table~\ref{tbl:abunds}.

We were unable to place any constraints on carbon in Tri~II. The GRACES spectra are not usable below 4800{\AA}, so the $G$-band cannot be measured.
The CH lists from \citet{Masseron14,Kurucz11} do suggest strong CH features should exist at 5893{\AA} and 8400{\AA} that were used to place a [C/Fe] upper limit by \citet{Venn17}, but we could not find these features in several carbon-enhanced metal-poor stars or atlas spectra of the Sun and Arcturus \citep{Hinkle03}\footnote{\url{ftp://ftp.noao.edu/catalogs/arcturusatlas}}.
No other C features are available.
\citet{Kirby17} were able to measure $\mbox{[C/Fe]} \sim -0.1$ for TriII-40 from their HIRES spectrum, so this star is not carbon enhanced.

\subsection{$\alpha$-elements: Mg, Si, Ca, Ti}
The abundances of these four $\alpha$-elements are determined from equivalent widths.
The magnesium abundance is determined from $3-6$ lines, but always using two of the Mg b lines.
Silicon can only be measured in the Gru~I stars, using the 4102{\AA} line that is in the wing of H$\delta$. The abundance uncertainty from only this single line is quite large, ${\approx}0.6$ dex.
Neutral calcium is well-determined by a large number of lines, and it should be considered the most reliable $\alpha$-element.
Titanium has several strong lines in both the neutral and singly ionized state, though only a handful ($1-5$) Ti lines can be measured in the Tri~II stars.
The abundance of Ti\,I is affected by NLTE effects \citep[e.g.,][]{Mashonkina17}, so we only plot Ti\,II abundances in Figure~\ref{f:grid1} both to avoid NLTE effects and because a Ti\,II line can be measured in all four of our stars. The literature sample also uses Ti\,II whenever possible.

\subsection{Odd-Z elements: Na, Al, K, Sc}
Sodium is measured from the Na D lines for GruI-032, GruI-038, and TriII-40.
While we can identify the presence of Na D lines in TriII-46, the lines are too noisy for a reliable abundance measurement. An upper limit $\mbox{[Na/Fe]} < 1.04$ is found from the subordinate Na lines near 8190\,{\AA}, and for completeness we include the best estimate of equivalent widths for the Na D lines in Table~\ref{tbl:eqw}.
NLTE corrections are not applied since most stars in the literature comparison sample do not have these corrections, but the grid from \citet{Lind11} gives corrections of $-0.28$ for GruI-032, $-0.32$ for GruI-038, and $-0.06$ for TriII-40.
The two Gru~I stars have solar ratios of Na, following the usual halo trend.
In contrast, TriII-40 has significantly subsolar $\mbox{[Na/Fe]} =-0.79 \pm 0.22$ that is an outlier from the halo trend, as first reported by \citet{Venn17}.
A similarly low [Na/Fe] ratio has previously been seen in one of three stars in the UFD Coma Berenices \citep{Frebel10b}. The primordial (first generation) population of stars in globular clusters also have low Na, but all with $\mbox{[Na/Fe]} > -0.5$, \citep{Gratton12}.

Aluminum and scandium are only measured in the Gru~I stars.
Al is determined from a single line at 3961{\AA}. Given the low S/N in this region, Al is the least certain abundance of all elements measured here. The measurement is consistent with that of other halo stars at $\mbox{[Fe/H]} \approx -2.5$, but it is not a meaningful constraint.
Sc lines in Gru~I are synthesized due to hyperfine structure \citep{Kurucz95}, and the abundances are also similar to other halo stars.
For completeness, we place Sc upper limits in the Tri~II stars with some weak red lines that provide no interesting constraint.

\begin{figure*}
\centering
\includegraphics[width=18cm]{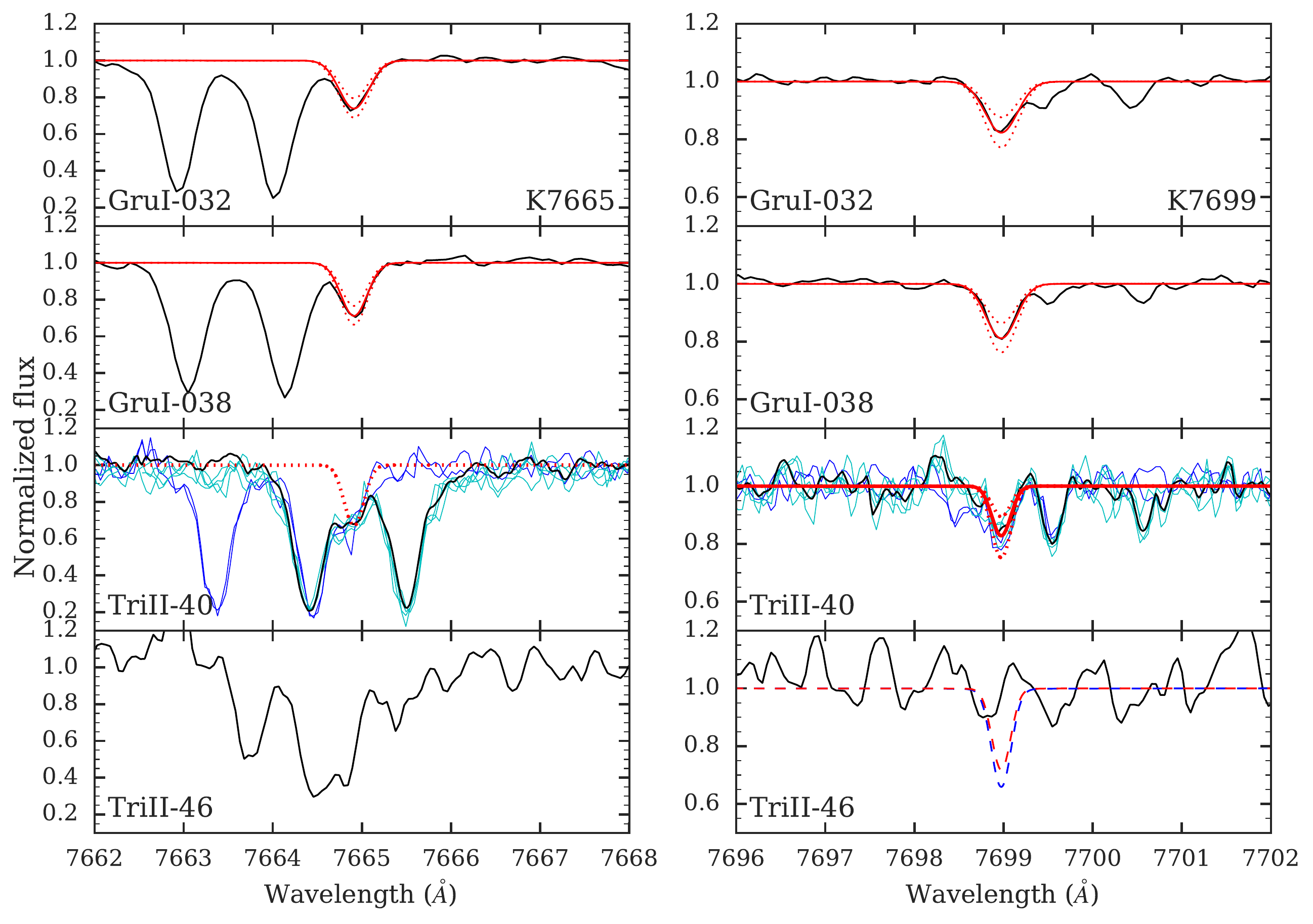}
\caption{Spectrum around K lines for our four stars.
Black lines are the data, solid red lines indicate synthesis fit, dotted red lines indicate uncertainty.
In the third row (TriII-40), dark blue lines indicate data from Dec 2015, while cyan lines indicate data from Sep 2016, showing how the location of telluric absorption shifts relative to the K line. The 7699 line is cleanly detected in Sep 2016 (telluric lines are at 7699.5 {\AA} and 7701.8 {\AA}). The same abundance is synthesized at the expected strength of the 7665 line, but we do not use that line because it is too blended with telluric absorption.
In the fourth row (TriII-46), the dashed red line is the K upper limit, and the dashed blue line indicates [K/Fe]=1.
\label{f:kspec}}
\end{figure*}

Potassium has two strong lines at 7665\,{\AA} and 7699\,{\AA}. These lines are located near several telluric absorption features.
Figure~\ref{f:kspec} shows these two lines and the best-fit synthetic spectrum or upper limits.
The top two spectra are Gru~I observations, where observations were conducted within the span of one month, so the telluric features do not move much due to heliocentric corrections. Both Gru~I stars have K lines that are easily distinguished from the telluric features.
The bottom two spectra of Figure~\ref{f:kspec} are Tri~II observations, which were conducted in Dec 2015 and Sep 2016.
The heliocentric correction is different between these epochs by ${\sim}40$ km/s, so the telluric features shift by ${\approx}1${\AA} between 2015 and 2016.
We emphasize this for TriII-40 by showing individual frames from Dec 2015 (thin blue lines) and Sep 2016 (thin cyan lines).
Note that we used \texttt{scombine} in IRAF with \texttt{avsigclip} rejection to obtain the coadded black spectra, so it tends to follow the telluric lines from Sep 2016 (four exposures) rather than Dec 2015 (two exposures).
The 7699\,{\AA} line is detected in TriII-40. It is significantly blended with a telluric line in the Dec 2015 observations (see \citealt{Venn17} figure 4, dark blue lines here), but cleanly separated in the Sep 2016 observations.
We find [K/Fe] = 0.8 in TriII-40, in agreement with the measurement by \citet{Venn17}.
The 7665\,{\AA} line is severely blended with telluric lines in both epochs, so we do not use it but just highlight its position in Figure~\ref{f:kspec} with a synthesized K line.
Neither K line is detected for TriII-46, and an upper limit $\mbox{[K/Fe]} < 0.77$ is set with the 7699\,{\AA} line.
We could not account for the telluric lines when setting this upper limit, but this makes the limit more conservative.
NLTE corrections have not been applied, but they can be large (as high as $-0.4$ dex for the most K-enhanced stars in LTE, \citealt{Andrievsky10}).

\begin{figure}
\centering
\includegraphics[width=8cm]{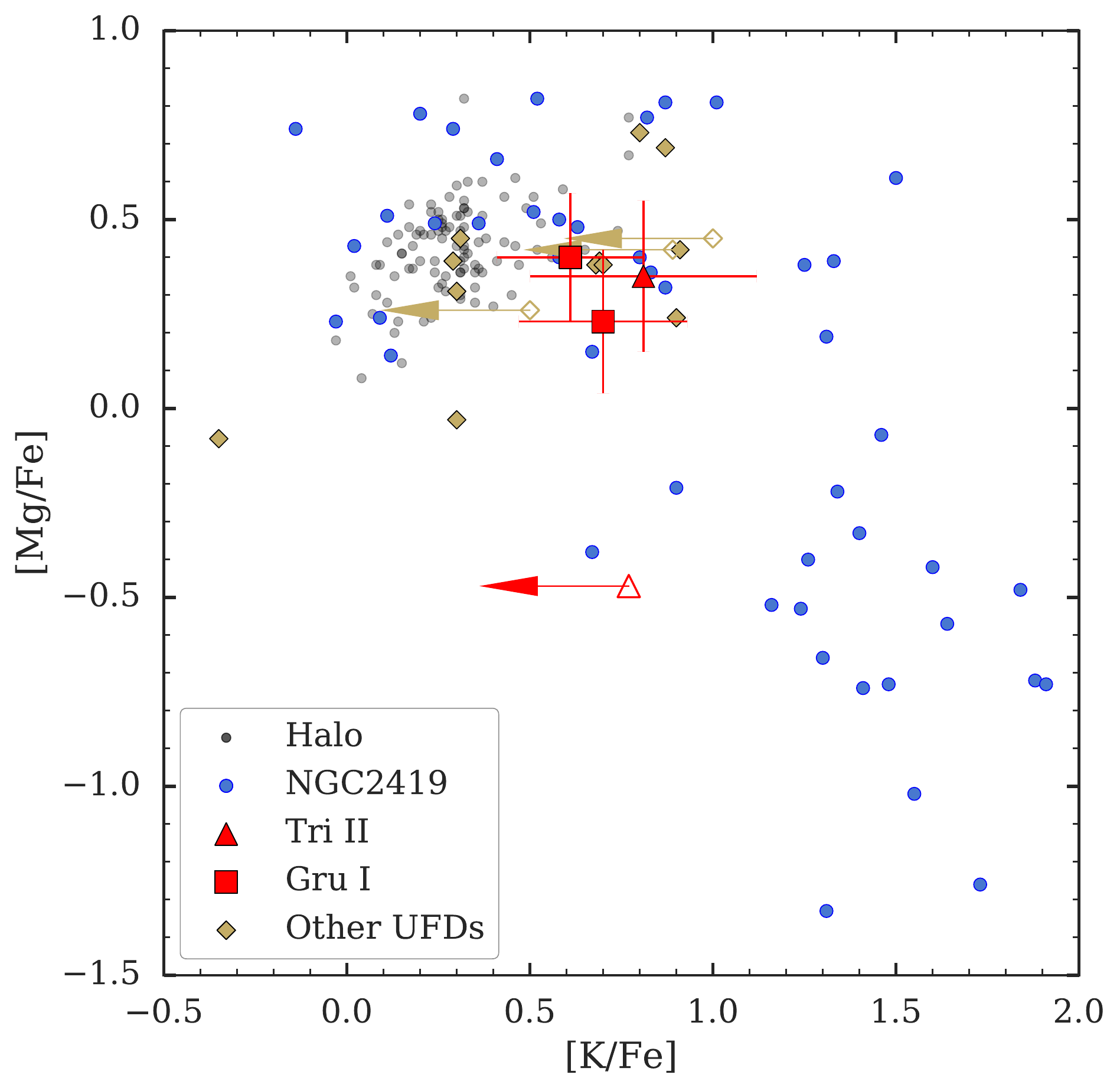}
\caption{K and Mg abundances of stars in the Tri~II and Gru~I from this work, compared to K and Mg in the stellar halo \citep{Roederer14c}, NGC2419 \citep{Mucciarelli12} and other UFDs (Boo~II, \citealt{Ji16a}; Ret~II, \citealt{Ji16c}; Tuc~II, \citealt{Ji16d}; Segue~2, \citealt{Roederer14a}; and Tuc~III, \citealt{Hansen17}).
The K abundance of TriII-46 is not enhanced, so Tri~II does not follow the strange K-Mg anticorrelation in NGC2419.
Note that the halo sample here is different than in Figures~\ref{f:grid1} and \ref{f:grid2} because our usual halo compilation does not have K abundances \citep{Frebel10}.
Adapted from \citet{Venn17}.
\label{f:kmg}}
\end{figure}

\subsection{Iron-peak elements: Cr, Mn, Co, Ni, Zn}
The Fe-peak abundances were determined with equivalent widths, except for Mn, which is synthesized due to hyperfine structure \citep{Kurucz95}.

In Gru~I we can constrain Cr, Mn, Co, and Ni, finding that both stars have essentially identical abundances of these elements.
Though the Cr and Mn abundances are similar to those in metal-poor stars in other UFDs or in the Milky Way halo, the Co and Ni abundances are somewhat higher.
However, this difference is not very significant, especially for Co which is derived from only a few bluer lines.
One Zn line is marginally detected in GruI-032 with an abundance consistent to the halo trend, though with large uncertainty.

In Tri~II, we can detect Cr and Ni in TriII-40 and provide upper limits in TriII-46. Mn, Co, and Zn are unconstrained as they only have strong lines blueward of 5000{\AA}.
The upper limits for Cr and Ni in TriII-46 are uninteresting.
For TriII-40, we detect a normal [Cr/Fe] ratio, but Ni appears significantly enhanced ($\mbox{[Ni/Fe]} = 0.57 \pm 0.16$), in agreement with \citet{Kirby17} and \citet{Venn17}.

\subsection{Neutron-capture elements: Sr, Ba, Eu}
\begin{figure*}
\centering
\includegraphics[width=18cm]{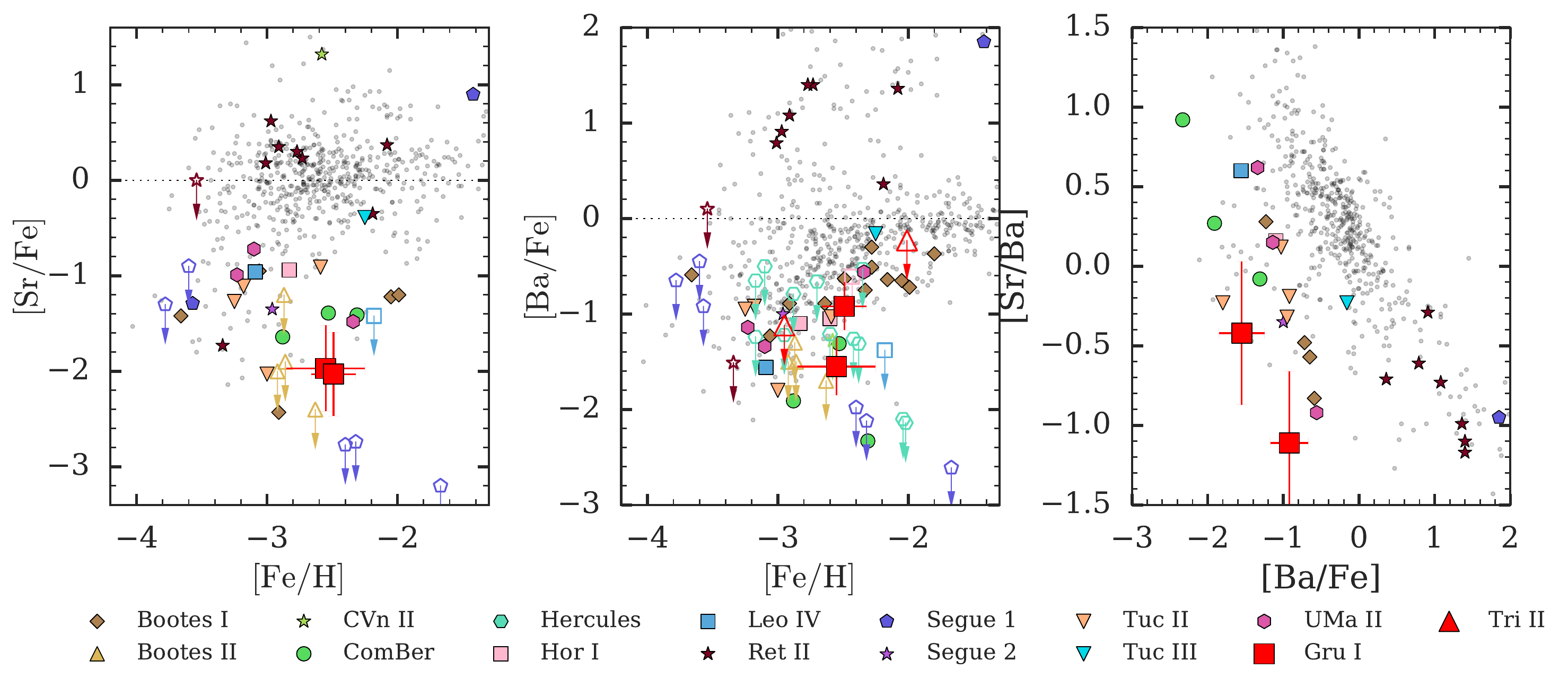}
\caption{Abundances of Sr and Ba in UFDs compared to halo stars.
Symbols are as in Figure~\ref{f:grid1}.
The left two panels show the abundance trend with respect to [Fe/H].
Note that there is no constraint on Sr for Tri~II stars.
The rightmost panel shows that most halo stars cluster near $\mbox{[Sr,Ba/Fe]} \approx 0$, but most UFDs are clearly offset to lower Sr and Ba.
\label{f:grid2}}
\end{figure*}

Strontium is detected only in Gru~I, as the strong Sr\,II lines at 4077{\AA} and 4215{\AA} are out of the range of the Tri~II (GRACES) spectra. The abundance of both lines is determined with spectrum synthesis.
The Sr abundances in these two stars are very similar, $\mbox{[Sr/Fe]} \approx -2$, which is much lower than what is found in most halo stars but similar to most UFDs (Figure~\ref{f:grid2}).

Barium is measured with four different lines in the Gru~I stars including hyperfine structure and isotope splitting \citep{McWilliam98}.
We use solar isotope ratios \citep{Sneden08}, but given the low overall abundance, changing this to $r$- or $s$-process ratios does not significantly affect our abundances.
GruI-032 has a low $\mbox{[Ba/Fe]} \approx -1.6$, but GruI-038 has a much higher Ba abundance $\mbox{[Ba/Fe]} \approx -1.0$.
This is formally only 1.6$\sigma$ different, but differential comparison of the line strengths (e.g., the 6497{\AA} line in Figure~\ref{f:spec}) suggests that the difference is real.
We discuss this more in Section~\ref{s:discoutlier}, but both Ba abundances are low and similar to those in most UFDs.

Ba is not detected in either Tri~II star, so instead we place $5\sigma$ upper limits. The Ba limit for TriII-40 is $\mbox{[Ba/Fe]} < -1.25$, suggesting a low Ba abundance similar to other UFDs.
\citet{Kirby17} determined $\mbox{[Sr/Fe]}=-1.5$ and $\mbox{[Ba/Fe]}=-2.4$ from their HIRES spectrum of this star, consistent with our upper limit and showing TriII-40 clearly has very low neutron-capture element abundances.
The Ba limit for TriII-46 is only $\mbox{[Ba/Fe]} \lesssim -0.2$, but this is still at the lower envelope of the halo trend (Figure~\ref{f:grid2}).

Eu is not detected in any of these four stars, as expected given the low Sr and Ba abundances. Upper limits are placed from the 4129{\AA} line for the Gru~I stars (MIKE data) and from the 6645{\AA} line for the Tri~II stars (GRACES data).

\section{Discussion}
\label{s:discussion}
\subsection{Abundance Anomalies}\label{s:discoutlier}
The abundance ratios of the two stars in Gru~I are nearly identical to each other, and similar to typical UFD stars at $\mbox{[Fe/H]} \approx -2.5$. The most notable exception is the Ba abundance, where GruI-038 has 0.6 dex higher [Ba/Fe] than GruI-032. After applying corrections from \citet{Placco14}, GruI-038 also has a higher carbon abundance than GruI-032 ($\mbox{[C/Fe]} = +0.57$ vs $+0.21$, respectively).
The differences are both somewhat low significance, and it is reasonable to consider these two stars chemically identical.
However if the differences are real, one possible explanation is that GruI-038 formed from gas that had been polluted by more AGB stars compared to GruI-032.
A lower mass ($1-4 M_\odot$) AGB star could add significant Ba and C without changing the Sr abundance too much \citep[e.g.,][]{Lugaro12}.
Since AGB winds are low velocity, their C and Ba production would be more inhomogeneously distributed in the star-forming gas of a UFD progenitor \citep[e.g.,][]{Emerick18}. 
However, many more stars in Gru~I would be needed to test this scenario.

We confirm the result from \citet{Venn17} that TriII-46, the more Fe-rich star in Tri~II, has very low [Mg/Fe] and [Ca/Fe] ratios.
The standard interpretation is that TriII-46 must have formed after significant enrichment by Type~Ia supernovae, and this star does follow the decreasing [$\alpha$/Fe] trend of other stars in Tri~II \citep{Kirby17}.
Indeed, most other UFDs show a similar downturn in [$\alpha$/Fe] ratios as [Fe/H] increases \citep{Vargas13}, though Horologium~I is unique in that all known stars in the system have low [$\alpha$/Fe] \citep{Nagasawa18}, and Segue~1 is unique in that it shows no downturn in $\alpha$-elements at high [Fe/H] \citep{Frebel14}.
It is actually somewhat surprising that the very low-luminosity Tri~II appears to have formed stars long enough to be enriched by Type~Ia supernovae, since its luminosity is very similar to Segue~1.
If Tri~II were significantly tidally stripped by now \citep{Kirby15,Kirby17,Martin16} this would help reconcile enrichment by Type~Ia supernovae with the small present-day luminosity.
However, the orbital pericenter of Tri~II is 20 kpc, where tidal effects are not too strong \citep{Simon18}; and there are no visible signs of tidal disruption in deep imaging \citep{Carlin17}.
An alternate explanation could be the presence of very prompt Type~Ia supernovae \citep[e.g.,][]{Mannucci06}.
If this is the case, it may have implications for the single-degenerate vs. double-degenerate debate of Type~Ia supernova progenitors.
Short detonation delay times (${\sim}100s$ of Myr) are a common feature of double-degenerate models, and less common (though still possible) for single-degenerate models \citep[e.g.,][]{Maoz14}.
One way to distinguish these models in Tri~II would be to examine Fe-peak elements like Mn, Co, and Ni \citep[see][]{McWilliam18}; but these elements are unavailable in our GRACES spectra.

\citet{Venn17} first noticed that the K and Mg abundances in Tri~II could match the unusual globular cluster NGC2419, which displays a K-Mg anticorrelation of unknown origin \citep{Cohen12,Mucciarelli12}.
If so, then TriII-46 should have very high $1 < \mbox{[K/Fe]} < 2$ (Figure~\ref{f:kmg}).
Our new limit of $\mbox{[K/Fe]} \lesssim 0.8$ in TriII-46 suggests that Tri~II probably does not display the same K-Mg anticorrelation as NGC2419.
[K/Fe] is often enhanced in LTE, both for UFD stars and halo stars \citep{Roederer14c}.
NLTE effects tend to amplify the strengths of the resonance lines for K-enhanced stars, so they likely contribute to the apparent overabundance of K in these stars \citep{Andrievsky10}.

We also confirm results from \citet{Kirby17} and \citet{Venn17} that TriII-40 has very low $\mbox{[Na/Fe]} = -0.79 \pm 0.22$ and somewhat high $\mbox{[Ni/Fe]} = 0.57 \pm 0.16$.
This star has $\mbox{[Fe/H]} \sim -3$ and enhanced $\alpha$-elements, so we would nominally expect its abundance ratios to predominantly reflect the yields of metal-poor core-collapse supernovae (CCSNe).
It is somewhat counterintuitive to find enhanced Ni and depressed Na in a CCSN, as the production of both elements is positively correlated with the neutron excess in a supernova \citep[e.g.,][]{Venn04,Nomoto13}.
However, this appears to break down at the lowest metallicities, and the online \emph{Starfit} tool\footnote{\url{http://starfit.org/}} finds that a Pop\,III supernova progenitor (11.3 $M_\odot$, $E=3 \times 10^{51}$\,erg, from the supernova yield grid of \citealt{Heger10}) provides a decent fit to the Mg, Ca, Ti, Fe, and Ni abundances ($\mbox{[Na/Fe]} \approx -1.0$, $\mbox{[Ni/Fe]} \approx +0.2$).
An alternate possibility is that this $\mbox{[Fe/H]} \sim -3$ star formed from gas already affected by Type~Ia supernovae, as Chandrasekhar mass explosions can produce high [Ni/Fe] \citep[e.g.,][]{Fink14} while reducing [Na/Fe] by adding iron. It seems very unlikely to form and explode a white dwarf so early in this galaxy's history, but age and metallicity may be decoupled at early times due to inhomogeneous metal mixing \citep[e.g.,][]{Frebel12,Leaman12,Nomoto13}. A very prompt population of Type Ia's with merging delay times as low as 30 Myr could also exist \citep{Mannucci06}.
We note that the Na and Ni lines in our spectrum of TriII-46 are very noisy and cannot provide a reliable abundance, but the best-fit abundance estimates (Table~\ref{tbl:eqw}) do suggest this star also has low [Na/Fe] and enhanced [Ni/Fe].

\subsection{Classification as dwarf galaxy or globular cluster} \label{s:discclassify}

In this paper, we consider three criteria that can be used to classify Tri~II and Gru~I as either ultra-faint dwarf galaxies or globular clusters.

\noindent
(1) a velocity dispersion indicating the presence of dark matter, \\
\noindent
(2) an [Fe/H] spread implying the ability to form multiple generations of stars despite supernova feedback, or significant internal mixing, and \\
\noindent
(3) unusually low neutron-capture element abundances compared to halo stars.

The first two criteria were codified by \citet{Willman12} and imply that the stellar system is the result of extended star formation in a dark matter halo.
The third criterion is based on previous studies of UFDs confirmed by the other two criteria \citep[e.g.,][]{Frebel10b, Frebel14, Frebel15, Simon10, Koch13}, and it has recently been used as a way to distinguish UFD stars from other stars \citep[e.g.,][]{Kirby17,Casey17,Roederer17}.
Unlike the first two criteria, this is a criterion specifically for the lowest mass galaxies, rather than defining galaxies in general.
Note that violating the criterion also does not preclude an object from being a UFD, as is evident from the $r$-process outliers Ret~II and Tuc~III that experienced rare $r$-process enrichment events. However, when multiple stars are observed in the same UFD, the majority of stars do tend to have similar neutron-capture element abundances.
We discuss possible explanations for criterion (3) in Section~\ref{s:discwhy}, but first accept it as an empirical criterion.

\subsubsection{Triangulum~II}
The case of Tri~II was already extensively discussed by \citet{Kirby15,Kirby17,Martin16,Venn17,Carlin17}, generally finding that it is most likely a UFD rather than a star cluster.
Our high-resolution abundance results are consistent with the discussion in \citet{Venn17} and \citet{Kirby17}, namely that we find a difference in [Fe/H] between these two stars at about $2\sigma$ significance, and TriII-46 has lower [$\alpha$/Fe] ratios compared to TriII-40.
\citet{Kirby17} previously found very low Sr and Ba abundances in TriII-40, and our Ba limit on TriII-46 is consistent with overall low neutron-capture element abundances in Tri~II (though additional data is needed to confirm that TriII-46 is well below the halo scatter).
Tri~II thus likely satisfies criteria (2) and (3), though it is unclear if it satisfies criterion (1) (see \citealt{Kirby17}, figure~2).
Our main additional contribution here is a more stringent upper limit on K in TriII-46
(Figure~\ref{f:kmg}) as discussed above in Section~\ref{s:discoutlier}, which shows Tri~II does not have the abundance signature found in the globular cluster NGC2419.

\subsubsection{Grus~I}
\citet{Walker16} identified seven probable members in Gru~I. This sample was insufficient to resolve either a velocity dispersion or metallicity dispersion.
Our high-resolution followup of two stars has found that those stars have indistinguishable [Fe/H].
Thus, Gru~I does not currently satisfy criteria (1) or (2) to be considered a galaxy.
However, we have found that the neutron-capture element abundances in Gru~I are both low and similar to UFDs, satisfying criterion (3). Gru~I thus most likely appears to be a UFD, and we expect that further spectroscopic study of Gru~I will reveal both metallicity and velocity dispersions.
We note that the velocity difference in our two Gru~I stars alone does already suggest a potentially significant velocity dispersion.

The mean metallicity determined by \citet{Walker16} for Gru~I is $\mbox{[Fe/H]} \sim -1.4 \pm 0.4$, which placed it far from the luminosity-metallicity trend of other dSph galaxies, while globular clusters do not have such a relationship.
However, the two brightest stars, analyzed here, both have $\mbox{[Fe/H]} \sim -2.5$ that would be consistent with the mean trend. 
Only ${\sim}0.3$\,dex of the difference can be attributed to their metallicity zero-point offset (see Section~\ref{s:litcomp}). The rest of the discrepancy is due to the fact that \citet{Walker16} found their other five members of Gru~I to have a much higher [Fe/H] than these two stars, ranging from [Fe/H] = $-2$ to $-1$.
Those five fainter stars are over 1 mag fainter than our stars, currently out of reach for high-resolution spectroscopic abundances so we cannot test the true metallicity of Gru~I with our data.
However, those stars also have very low S/N, and inferred effective temperatures that are much higher than expected based on photometry alone.
We thus suggest the metallicity of Gru~I is probably closer to the value measured from our two stars.
Recently, \citet{Jerjen18} published deep photometry of Gru~I, with isochrone-based metallicities of $\mbox{[Fe/H]} = -2.5 \pm 0.3$.

\subsection{Why do most UFDs have low neutron-capture element abundances?}\label{s:discwhy}
\begin{figure*}
\includegraphics[width=18cm]{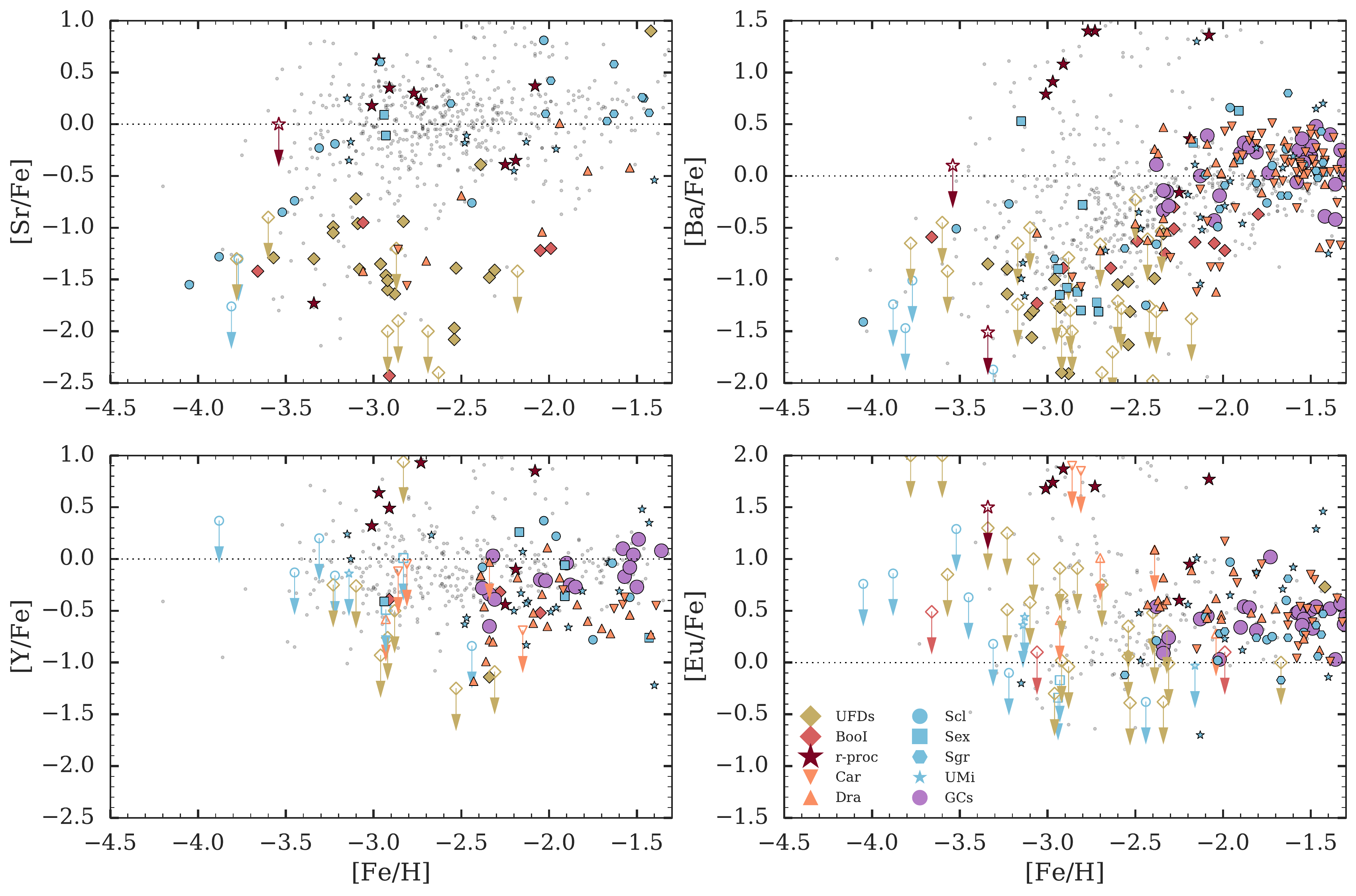}
\caption{Neutron-capture element abundances for UFDs (yellow diamonds; separating Boo~I as red diamonds; and Ret~II and Tuc~III as large dark red stars), classical dSphs (blue and orange symbols), globular clusters (large purple circles), and halo stars (grey points).
Classical dSph stars come from \citealt{Aoki09,Cohen09,Cohen10,Frebel10a,Fulbright04,Geisler05,JHansen18,Jablonka15,Kirby12,Norris17b,Shetrone01,Shetrone03,Simon15Scl,Skuladottir15,Tafelmeyer10,Tsujimoto15a,Tsujimoto17,Ural15,Venn12}.
\label{f:ncapcomp}}
\end{figure*}

\begin{figure}
\includegraphics[width=8.5cm]{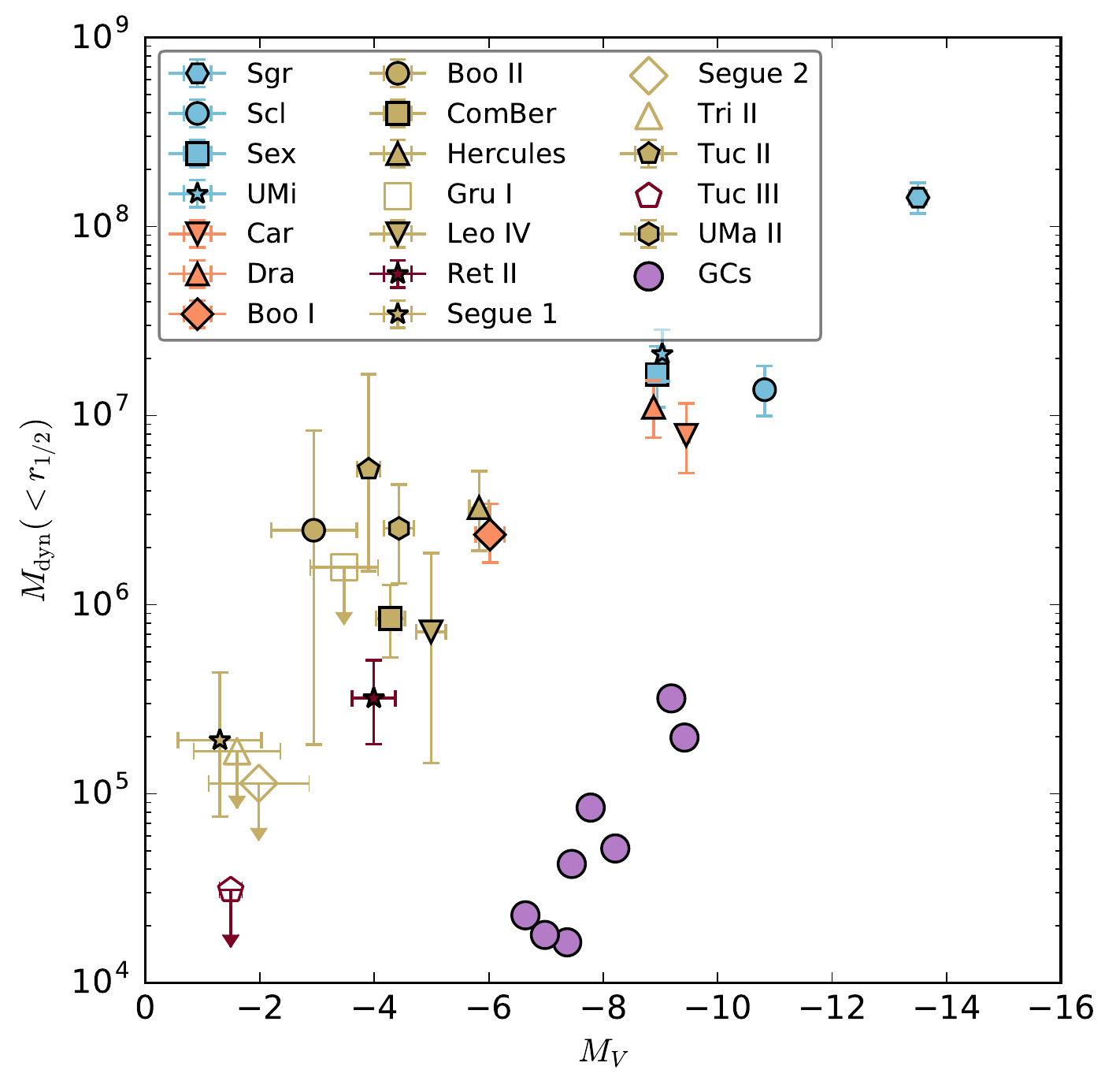}
\caption{Absolute $V$ magnitude vs dynamical mass within half light radius for dSphs with neutron-capture element constraints.
Galaxies are color-coded according to their [Sr/Fe] and [Ba/Fe] abundance at $-3.5 \lesssim \mbox{[Fe/H]} \lesssim -2.5$.
Yellow points have both low Sr and Ba, orange points have low Sr but regular Ba, blue points have regular Sr and Ba.
For comparison, we also show globular clusters in the \citet{Pritzl05} sample with $\mbox{[Fe/H]} \lesssim -2$.
The dynamical data and luminosity for dwarf galaxies come from \citet{Munoz18}, supplemented by \citealt{Majewski03,Bechtol15}.
Velocity dispersions are from \citealt{Bellazzini08,Kirby13a,Kirby17,Koch09,Koposov11,Simon07,Simon11,Simon15,Simon17,Simon19,Walker09c,Walker09,Walker16}.
$M_{\rm dyn}$ is computed with the equation in \citet{Walker09}.
Globular cluster data are from \citet{Harris10}.
\label{f:dsphbyncap}}
\end{figure}

Figure~\ref{f:ncapcomp} shows the neutron-capture element abundances of UFD stars relative to halo stars and classical dSph stars.
Excluding Ret~II and Tuc~III, it is clear that UFDs have low neutron-capture element abundances relative to these other populations in both Sr and Ba, and most apparent in Sr.
The astrophysical origin (or origins) of these low-but-nonzero neutron-capture element abundances is still an open question (see Section~\ref{s:discncap}).

However, the abundance signature of this (or these) low-yield site(s) is usually hidden in more metal-rich stars.
This is clearly seen by examining the classical dSph galaxies, which are somewhat more evolved than UFDs due to their higher mass.
In Sculptor, we can see a ${>}1$ dex rise in [Sr/Fe] from UFD levels to typical halo star levels occurring at very low metallicity $-4 < \mbox{[Fe/H]} < -3.3$; while a rise in [Ba/Fe] occurs later, at $\mbox{[Fe/H]} \sim -2.5$ \citep[also see][]{Jablonka15,Mashonkina17}.
Similar trends exist for Sagittarius, Sextans, and Ursa Minor.
We highlight Draco and Carina separately, as their stars' [Sr/Fe] ratios stay similarly low to UFDs until $\mbox{[Fe/H]} \gtrsim -2.5$, but unlike UFDs their [Ba/Fe] ratios rise with [Fe/H].
The UFD Boo~I is similar to Draco and Carina and unlike most UFDs in this sense, as well.
The rise in Sr and Ba suggests the delayed onset of different, more prolific, sources of neutron-capture elements, presumably some combination of AGB stars and neutron star mergers.
These higher-yield later-onset sources of Sr and Ba will eventually dominate total Sr and Ba production.
Overall, it seems that larger galaxies manage to reach a ``normal'' halo-like neutron-capture element abundance at lower [Fe/H] than smaller galaxies, implying that they can be enriched by those dominant sources of Sr and Ba \citep[also see][]{Tafelmeyer10,Jablonka15}.

The question of why UFDs have low neutron-capture element abundances thus boils down to why these high-yield sources of neutron-capture elements do not contribute metals to most UFDs while they are forming stars.
We can imagine three possible reasons:
\begin{enumerate}
\item UFDs do not form enough stars to fully sample all metal yields from a stellar population.
If the dominant sources of Sr and Ba are produced rarely or stochastically, they will only occasionally enrich a given UFD; so most UFDs would have low [Sr,Ba/Fe] \citep[e.g.,][]{Koch08,Koch13,Simon10,Venn12,Venn17,Ji16b}.
\item UFDs form in small potential wells, so they do not retain metals very well \citep[e.g.,][]{Kirby11outflow,Venn12}. If the dominant sources of Sr and Ba are lost with higher efficiency in UFDs (relative to iron), this would result in low [Sr,Ba/Fe].
\item UFDs form stars for only a short time. If the dominant neutron-capture element sources have long delay times (e.g., neutron star mergers or AGBs), these sources may only produce metals after UFDs have finished forming stars. Then, surviving UFD stars would not preserve the metals from those sources.
\end{enumerate}
We note that Sr and Ba appear to have differing trends, so the explanations for Sr and Ba may differ as well.

As one attempt to distinguish between these possibilities, we consider whether there are correlations with stellar mass or current dynamical mass.
Figure~\ref{f:dsphbyncap} shows the absolute magnitude and inferred dynamical mass within the half light radius for several classical dSphs and UFDs.
The yellow points are UFDs that have low [Sr/Fe] and [Ba/Fe]. 
Blue points are classical dSphs (UMi, Sex, Scl, Sgr) that have regular Sr and Ba trends. 
In orange we highlight Boo~I, Carina, and Draco, which have low Sr at $\mbox{[Fe/H]} \sim -3$, but Ba behavior similar to the more massive dSphs. 
We also note that Draco, UMi, and all UFDs have CMDs indicating purely old stellar populations ($>10-12$\,Gyr old), while more luminous dSphs (Carina and above) show evidence for some late time star formation \citep{Weisz14a,Brown14}.
There is a broad transition in neutron-capture element content occurring somewhere between $-6 > M_V > -10$ and $10^6 < M_{\rm dyn}/M_\odot < 10^7$, also roughly corresponding to the purely old dSphs.
Unfortunately, given the strong correlations between luminosity, dynamical mass, and overall age in this sample, it is hard to distinguish between the three reasons listed above for low neutron-capture elements in UFDs.

Explanation (2) is somewhat disfavored if one accepts two stronger assumptions.
First, $M_{\rm dyn}(<r_{1/2})$ is not a good measure of the total halo mass, because the half light radius is only a tiny fraction of the overall halo size. Correcting for this requires extrapolating an assumed density profile to larger radii, but such extrapolations imply that UFDs and even some of the larger dSphs may all reside in dark halos of similar mass \citep{Strigari08}. A similar dark halo mass is also expected from a stellar-mass-to-halo-mass relation with large intrinsic scatter \citep[e.g.,][]{Jethwa18}.
Second, one must assume that $z=0$ halo masses are highly correlated with halo masses at the time of star formation. This is true on average in $\Lambda$CDM, but it  breaks down in specific cases due to scatter in halo growth histories \citep[e.g.,][]{Torrey15} and tidal stripping from different subhalo infall times \citep[e.g.,][]{Dooley14}.
Together, these two assumptions would imply that neutron-capture element behavior is uncorrelated with halo mass, disfavoring explanation (2).
Furthermore, comparison to classical dSphs suggests the short star formation timescale (3) is unlikely for Sr: more massive dSphs like Scl and Sgr are much more efficient at forming stars, but they are already Sr-enriched at $\mbox{[Fe/H]} \sim -3$.
It may thus be the case that explanation (1) is the most likely one for Sr, i.e. that the dominant source of Sr is stochastically produced.
However, explanations (1) and (3) both remain viable for Ba; and explanation (2) remains for both Sr and Ba as well if the two stronger assumptions do not hold.

\subsection{Comparison to globular clusters}\label{s:gc}
Globular clusters (GCs) have very different neutron-capture element abundances than UFDs.
Figure~\ref{f:ncapcomp} shows the mean abundances of GCs as purple circles (compiled in \citealt{Pritzl05}\footnote{We have removed NGC 5897, NGC 6352, and NGC 6362 from this compilation, which were outliers in [Ba/Fe]. These three GCs were all observed by \citet{Gratton87} and scaled to a common $\log gf$ scale by \citet{Pritzl05}. However, the abundances derived by \citet{Gratton87} appear to conflict with the $\log gf$, and we suspect a typographical error for $\log gf$. We confirm this in NGC 5897 with more recent measurements by \citet{Koch14c}.}).
Sr is usually not measured in GCs, so we also show Y (which has similar nucleosynthetic origins as Sr).
It is immediately obvious that all neutron-capture elements in globular clusters closely trace the overall halo trend, as well as more metal-rich stars in classical dSphs.
In contrast, UFDs tend to lie at the extremes of the halo trend.

The origin of globular clusters is unknown, but one class of theories posits that metal-poor GCs form as the dominant stellar component of a small dark matter halo, rather than as a part of a larger galaxy \citep[e.g.,][and references therein]{Forbes18}.
Such theories usually have GCs form in the \emph{same} dark matter halos as UFDs (i.e., ${\sim}10^8 M_\odot$ dark matter halos that experience atomic line cooling), but something (e.g., a gas-rich merger) triggers them to become GCs instead of UFDs \citep[e.g.,][]{Griffen10,Trenti15,Ricotti16,Creasey18}.
However, if GCs do form in these small atomic cooling halos, their neutron-capture element enrichment should match that of UFDs, i.e. be very low, or at least show significant GC-to-GC scatter\footnote{At least one metal-poor globular cluster, M15, does show a significant \emph{internal} dispersion in neutron-capture element abundances ($>0.6$\,dex; \citealt{Sneden97}).
Some other GCs might also display such a dispersion, though it is much smaller ($0.3$\,dex) and could be due to systematic effects \citep{Roederer11b,Cohen11b,Roederer15}.
Either way, this dispersion is not enough to match the neutron-capture element deficiency seen in most UFD stars with $\mbox{[Fe/H]} \gtrsim -2.5$.}.
The difference in neutron-capture element abundances thus seems to imply that the known metal-poor GCs in the Milky Way formed as part of larger galaxies (e.g., \citealt{BoylanKolchin17}), rather than in their own dark matter halos.

Note that the neutron-capture element abundances are not affected by the multiple abundance populations usually discussed in globular clusters \citep[e.g.,][]{Gratton04,Roederer11b}.
Those variations in lighter abundances are due to an internal mechanism, rather than tracing the natal abundance of the gas the GCs formed from \citep[see e.g.,][and references therein.]{Gratton12,Bastian18}.

\subsection{On the origin of the ubiquitous neutron-capture element floor} \label{s:discncap}
We briefly discuss the most viable candidates for this ubiquitous low-yield neutron-capture element source occurring at low metallicity.
This is important not just for understanding UFD enrichment, but also for the most metal-poor halo stars, where Sr and/or Ba appear to be ubiquitously present at the level $\mbox{[Sr,Ba/H]} \sim -6$ 
\citep{Roederer13}\footnote{To our knowledge, the only star with limits below this threshold is a star with no detected Fe, SMSS 0313$-$6708, with extremely low limits $\mbox{[Sr/H]} < -6.7$ and $\mbox{[Ba/H]} < -6.1$ \citep{Keller14}.}.
The sources must explain the ubiquitous presence of both Sr and Ba, the overall low but nonzero yield of both Sr and Ba, and the fact that the [Sr/Ba] ratio in UFDs varies over ${\sim}2$ dex.

\emph{Neutrino-driven wind.}
The high-entropy neutrino-driven wind in CCSNe was initially thought to be a promising site for Sr and Ba production in the $r$-process \citep[e.g.,][]{Woosley92}, but contemporary simulations suggest wind entropies an order of magnitude too low to produce the full set of $r$-process elements up to uranium \citep[e.g.,][]{Arcones07}.
It still seems that this mechanism robustly produces a limited form of the $r$-process that always synthesizes Sr, but a little bit of Ba only under extreme conditions (e.g., neutron star mass $>2 M_\odot$, \citealt{Wanajo13}).
Supporting this, \citet{Mashonkina17} recently argued for two types of Sr production, one of which was highly correlated with Mg, implying CCSNe could produce Sr alone.
However, current models suggest that even extreme neutrino-driven winds cannot produce $\mbox{[Sr/Ba]} \sim 0$ \citep{Arcones11,Wanajo13}, so while they may be an important factor they probably are not the only source of neutron-capture elements in most UFDs.

\emph{Magnetorotationally driven jets.}
A dying massive star with extremely strong magnetic fields and fast rotation speeds can launch a neutron-rich jet that synthesizes copious Sr and Ba in the r-process \citep[e.g.,][]{Winteler12,Nishimura15}.
It is still debated whether such extreme conditions can be physically achieved in massive star evolution \citep[e.g.,][]{Rembiasz16a,Rembiasz16b, Mosta17}.
However, if the conditions are less extreme, such supernovae can actually produce both Sr and Ba without synthesizing the heaviest r-process elements in a delayed jet \citep{Nishimura15,Nishimura17,Mosta17}.
These more moderate rotation speeds and magnetic fields may be more plausible results of massive and metal-poor stellar evolution, and so the rate of these moderate jet explosions could occur much more often than is invoked to explain prolific $r$-process yields.
If so, then we propose that delayed magnetorotationally driven jets are a viable source of the low Sr and Ba abundances in UFDs.
Additional modeling focusing on the frequency of less-extreme jets is needed for a more detailed evaluation, and zinc abundances may help as well \citep{Ji18}.

\emph{Spinstars.}
Spinstars are rapidly rotating massive stars that can produce Sr and Ba in the $s$-process \citep[e.g.,][]{Meynet06}.
The amount of rotation changes the amount of internal mixing in the star, allowing these models to produce a wide range of [Sr/Ba] ratios,
though the amount of Ba is still subject to nuclear reaction rate uncertainties \citep{Cescutti13, Frischknecht16, Choplin18}.
The fiducial spinstar models in \citet{Frischknecht16} underproduce Sr and Ba by a factor of ${>}100$ to explain the observed values in UFDs \citep[e.g.,][]{Ji16d}, and having hundreds of spinstars in each UFD is unlikely given there are only hundreds of massive stars to begin with in each galaxy.
However, extreme spinstar models with particularly fast rotation velocities and a modified nuclear reaction rate increase the abundance yields by a factor $>10$ \citep{Cescutti13,Frischknecht12,Frischknecht16}\footnote{Yields from \url{http://www.astro.keele.ac.uk/shyne/datasets/s-process-yields-from-frischknecht-et-al-12-15}}.
These models also produce [C/Sr] and [C/Ba] $\sim +2.0$, consistent or somewhat lower than the C abundances in UFDs like Gru~I.
The [C/Fe] ratios are very high ($>3.0$), but the spinstar yields do not include any carbon or iron generated in a supernova explosion, which would reduce this extreme abundance ratio.
Thus, the extreme spinstar models are also a viable source for the neutron-capture elements found in UFDs.

Note that rotation is not the only way that neutron-capture processes can occur in metal-poor or metal-free stars, as it is just one of many possible mechanisms that can induce internal mixing and thus create free neutrons. Recently, \citet{Banerjee18a} and \citet{Clarkson18} have shown that proton ingestion into convective He shells can result in a low level of $s$-, $i$-, and $r$-processes in even in metal-free stars.
Some of the metal-poor models by \citet{Banerjee18a} are able to produce explain the low but nonzero amounts of Sr and Ba found in UFDs, as well as the diversity of [Sr/Ba] ratios.

\emph{An unknown low-yield r-process source.}
As of now, binary neutron star mergers are the only confirmed source of the full $r$-process (i.e., produces all elements from the 1st through 3rd $r$-process peaks).
However, there is evidence from halo stars with low Sr and Ba that UFDs are enriched by a low-yield (or heavily diluted) version of the same abundance pattern.
\citet{Roederer17} found three halo stars with low Sr and Ba as well as Eu detections consistent with the $r$-process ($-4 < \mbox{[Eu/H]} < -3.5$).
\citet{Casey17} found a halo star with $\mbox{[Sr,Ba/H]} \approx -6$, with $\mbox{[Sr/Ba]} \sim 0$ consistent with the full $r$-process.
Assuming that these halo stars originated in now-tidally-disrupted UFDs, that might imply that a low-yield but robust $r$-process does occur.
This has long been assumed to take place in some subset of core-collapse supernovae, but as mentioned above, current models cannot achieve this reliably.
However, UFDs display variations in [Sr/Ba] that cannot be explained by just a single $r$-process.

Disentangling these different sites will require determining abundances of neutron-capture elements other than Sr and Ba in UFD stars. Given the distance to known UFDs, this will require significant time investments with echelle spectrographs on 30m class telescopes.
In the meantime, progress can be made by study of bright, nearby halo stars with low Sr and Ba abundances \citep[e.g.,][]{Roederer17}.
For this purpose, the best stars are the relatively Fe-rich but Sr- and Ba-poor stars, as these are the ones most clearly associated with UFDs (Figure~\ref{f:ncapcomp}).
Such stars are expected to comprise $1-3$\% of halo stars at $-2.5 < \mbox{[Fe/H]} < -2.0$ \citep{Brauer18}.

\section{Conclusion} \label{s:conclusion}

We present detailed chemical abundances from high-resolution spectroscopy of two stars in Gru~I and two stars in Tri~II.
Overall, the abundance ratios of these stars are generally similar to those found in other ultra-faint dwarf galaxies, including extremely low neutron-capture element abundances.
The Gru~I stars are nearly chemically identical, except for possibly a different Ba abundance.
A possible similarity between Tri~II and the cluster NGC~2419 is probably ruled out by a new K upper limit, and there may also be an anomaly in Na and Ni (Section~\ref{s:discoutlier}).

The velocity and metallicity dispersions of Gru~I and Tri~II have not been decisive about whether they are ultra-faint dwarf galaxies or globular clusters, but we conclude they are both likely UFDs rather than GCs because both systems have extremely low neutron-capture element abundances (Section~\ref{s:discclassify}).
We thus expect future observations of these systems to confirm metallicity spreads, as well as significant velocity dispersions or signs of tidal disruption.

The low neutron-capture element abundances in UFDs reflect chemical enrichment at the the extreme low-mass end of galaxy formation in $\Lambda$CDM (Section~\ref{s:discwhy}): stochastic enrichment, metal loss in winds, and short star formation durations.
The dissimilarity in neutron-capture elements also suggests that globular clusters and UFDs do not form in the same environments, and thus that globular clusters probably did not form in their own dark matter halos (Section~\ref{s:gc}).
However, the nucleosynthetic origin of the low neutron-capture element abundances in UFDs like Gru~I and Tri~II is still an open question (Section~\ref{s:discncap}).

\acknowledgments
We thank Nidia Morrell for assisting with MIKE observations of Gru~I;
Kristin Chiboucas and Lison Malo for assistance with GRACES and data reduction;
Vini Placco for computing carbon corrections;
and Projjwal Banerjee, Gabriele Cescutti, Anirudh Chiti, Brendan Griffen, Evan Kirby, Andrew McWilliam, Tony Piro, and \'Asa Sk\'ulad\'ottir for useful discussions.
A.P.J. is supported by NASA through Hubble Fellowship grant HST-HF2-51393.001 awarded by the Space Telescope Science Institute, which is operated by the Association of Universities for Research in Astronomy, Inc., for NASA, under contract NAS5-26555.
J.D.S. and T.T.H. acknowledge support from the National Science Foundation under grant AST-1714873.
A.F. acknowledges support from NSF grants AST-1255160 and AST-1716251. 
K.A.V. acknowledges funding from the National Science and Engineering Research Council of Canada (NSERC), funding reference number 327292-2006. 
Based on observations obtained with ESPaDOnS, located at the Canada-France-Hawaii Telescope (CFHT). CFHT is operated by the National Research Council of Canada, the Institut National des Sciences de l'Univers of the Centre National de la Recherche Scientique of France, and the University of Hawai'i. ESPaDOnS is a collaborative project funded by France (CNRS, MENESR, OMP, LATT), Canada (NSERC), CFHT and ESA. ESPaDOnS was remotely controlled from the Gemini Observatory, which is operated by the Association of Universities for Research in Astronomy, Inc., under a cooperative agreement with the NSF on behalf of the Gemini partnership: the National Science Foundation (United States), the National Research Council (Canada), CONICYT (Chile), Ministerio de Ciencia, Tecnología e Innovación Productiva (Argentina) and Ministério da Ciência, Tecnologia e Inovação (Brazil).
This research has made use of the SIMBAD database, operated at CDS, Strasbourg, France \citep{Simbad},
and NASA's Astrophysics Data System Bibliographic Services.

\facilities{Magellan-Clay (MIKE, \citealt{Bernstein03}), Gemini-N (GRACES, \citealt{Chene14,Donati03})}
\software{CarPy \citep{Kelson03}, OPERA \citep{Martioli12}, IRAF, MOOG \citep{Sneden73,Sobeck11}, SMH \citep{Casey14}, \texttt{numpy} \citep{numpy}, \texttt{scipy} \citep{scipy}, \texttt{matplotlib} \citep{matplotlib}, \texttt{pandas} \citep{pandas}, \texttt{seaborn}, \citep{seaborn}, \texttt{astropy} \citep{astropy}}

\startlongtable
\begin{deluxetable*}{l|rr|rr|rr|rr|r|r}
\tablecolumns{11}
\tabletypesize{\footnotesize}
\tablecaption{Stellar Parameter Abundance Uncertainties\label{tbl:spabunderr}}
\tablehead{
 \colhead{} & \multicolumn{2}{c}{$\Delta\Teff$ (K)} & \multicolumn{2}{c}{$\Delta\logg$ (cgs)} & \multicolumn{2}{c}{$\Delta\nu_t$ (km/s)} & \multicolumn{2}{c}{$\Delta\mbox{[Fe/H]}$ (dex)} & \colhead{$\sigma_{\mbox{[X/H]}}$} & \colhead{$\sigma_{\mbox{[X/Fe]}}$}
}
\startdata
\hline
GruI-032 & $+155$ & $-155$ & $+0.37$ & $-0.37$ & $+0.32$ & $-0.32$ & $+0.19$ & $-0.19$ & $$ & $$ \\
\hline
$\mbox{[C/H]    }$ & $+0.31$ & $-0.22$ & $-0.09$ & $+0.11$ & $-0.01$ & $+0.01$ & $+0.10$ & $-0.09$ & 0.34 & 0.19 \\
$\mbox{[Na I/H] }$ & $+0.33$ & $-0.36$ & $-0.08$ & $+0.08$ & $-0.17$ & $+0.17$ & $-0.07$ & $+0.06$ & 0.41 & 0.14 \\
$\mbox{[Mg I/H] }$ & $+0.27$ & $-0.25$ & $-0.11$ & $+0.12$ & $-0.09$ & $+0.10$ & $-0.04$ & $+0.03$ & 0.31 & 0.05 \\
$\mbox{[Al I/H] }$ & $+0.32$ & $-0.36$ & $-0.12$ & $+0.12$ & $-0.19$ & $+0.19$ & $-0.06$ & $+0.05$ & 0.43 & 0.16 \\
$\mbox{[Si I/H] }$ & $+0.24$ & $-0.16$ & $-0.06$ & $+0.07$ & $-0.13$ & $+0.17$ & $-0.04$ & $+0.04$ & 0.30 & 0.11 \\
$\mbox{[K I/H]  }$ & $+0.22$ & $-0.21$ & $-0.05$ & $+0.06$ & $-0.06$ & $+0.07$ & $-0.04$ & $+0.03$ & 0.24 & 0.06 \\
$\mbox{[Ca I/H] }$ & $+0.17$ & $-0.17$ & $-0.06$ & $+0.06$ & $-0.05$ & $+0.06$ & $-0.04$ & $+0.03$ & 0.19 & 0.10 \\
$\mbox{[Sc II/H]}$ & $+0.08$ & $-0.05$ & $+0.09$ & $-0.08$ & $-0.14$ & $+0.16$ & $+0.03$ & $-0.02$ & 0.20 & 0.17 \\
$\mbox{[Ti I/H] }$ & $+0.28$ & $-0.35$ & $-0.07$ & $+0.09$ & $-0.05$ & $+0.07$ & $-0.05$ & $+0.04$ & 0.37 & 0.11 \\
$\mbox{[Ti II/H]}$ & $+0.07$ & $+0.02$ & $+0.11$ & $-0.11$ & $-0.10$ & $+0.12$ & $+0.03$ & $-0.02$ & 0.18 & 0.09 \\
$\mbox{[V I/H]  }$ & $+0.27$ & $-0.34$ & $-0.06$ & $+0.08$ & $-0.03$ & $+0.04$ & $-0.04$ & $+0.04$ & 0.35 & 0.11 \\
$\mbox{[Cr I/H] }$ & $+0.25$ & $-0.30$ & $-0.07$ & $+0.08$ & $-0.05$ & $+0.07$ & $-0.05$ & $+0.04$ & 0.32 & 0.06 \\
$\mbox{[Cr II/H]}$ & $-0.02$ & $+0.10$ & $+0.13$ & $-0.13$ & $-0.02$ & $+0.03$ & $+0.02$ & $-0.01$ & 0.17 & 0.10 \\
$\mbox{[Mn I/H] }$ & $+0.32$ & $-0.32$ & $-0.05$ & $+0.07$ & $-0.05$ & $+0.08$ & $-0.02$ & $+0.02$ & 0.34 & 0.09 \\
$\mbox{[Fe I/H] }$ & $+0.26$ & $-0.25$ & $-0.06$ & $+0.07$ & $-0.09$ & $+0.11$ & $-0.05$ & $+0.04$ & 0.30 & \nodata \\
$\mbox{[Fe II/H]}$ & $+0.01$ & $+0.11$ & $+0.12$ & $-0.12$ & $-0.10$ & $+0.12$ & $+0.03$ & $-0.02$ & 0.20 & \nodata \\
$\mbox{[Co I/H] }$ & $+0.37$ & $-0.33$ & $-0.07$ & $+0.09$ & $-0.20$ & $+0.24$ & $-0.06$ & $+0.06$ & 0.45 & 0.17 \\
$\mbox{[Ni I/H] }$ & $+0.25$ & $-0.25$ & $-0.04$ & $+0.06$ & $-0.05$ & $+0.06$ & $-0.04$ & $+0.03$ & 0.27 & 0.06 \\
$\mbox{[Zn I/H] }$ & $+0.05$ & $+0.00$ & $+0.07$ & $-0.06$ & $-0.01$ & $+0.01$ & $+0.01$ & $-0.01$ & 0.09 & 0.30 \\
$\mbox{[Sr II/H]}$ & $+0.01$ & $-0.03$ & $+0.07$ & $-0.10$ & $-0.15$ & $+0.17$ & $-0.01$ & $-0.02$ & 0.20 & 0.16 \\
$\mbox{[Ba II/H]}$ & $+0.11$ & $-0.08$ & $+0.12$ & $-0.10$ & $-0.02$ & $+0.03$ & $+0.03$ & $-0.02$ & 0.17 & 0.21 \\
\hline
GruI-038 & $+158$ & $-158$ & $+0.39$ & $-0.39$ & $+0.32$ & $-0.32$ & $+0.24$ & $-0.24$ & $$ & $$ \\
\hline
$\mbox{[C/H]    }$ & $+0.31$ & $-0.30$ & $-0.15$ & $+0.11$ & $-0.01$ & $+0.01$ & $+0.09$ & $-0.13$ & 0.37 & 0.23 \\
$\mbox{[Na I/H] }$ & $+0.21$ & $-0.31$ & $-0.08$ & $+0.04$ & $-0.17$ & $+0.18$ & $-0.06$ & $+0.02$ & 0.37 & 0.09 \\
$\mbox{[Mg I/H] }$ & $+0.19$ & $-0.26$ & $-0.12$ & $+0.08$ & $-0.08$ & $+0.08$ & $-0.03$ & $+0.00$ & 0.30 & 0.09 \\
$\mbox{[Al I/H] }$ & $+0.22$ & $-0.30$ & $-0.13$ & $+0.09$ & $-0.18$ & $+0.18$ & $-0.04$ & $+0.02$ & 0.38 & 0.12 \\
$\mbox{[Si I/H] }$ & $+0.19$ & $-0.22$ & $-0.04$ & $+0.02$ & $-0.14$ & $+0.17$ & $-0.03$ & $+0.02$ & 0.28 & 0.08 \\
$\mbox{[K I/H]  }$ & $+0.17$ & $-0.23$ & $-0.04$ & $+0.02$ & $-0.08$ & $+0.10$ & $-0.03$ & $+0.02$ & 0.26 & 0.06 \\
$\mbox{[Ca I/H] }$ & $+0.14$ & $-0.18$ & $-0.04$ & $+0.02$ & $-0.04$ & $+0.05$ & $-0.02$ & $+0.01$ & 0.19 & 0.12 \\
$\mbox{[Sc II/H]}$ & $+0.11$ & $-0.11$ & $+0.06$ & $-0.09$ & $-0.16$ & $+0.15$ & $+0.01$ & $-0.06$ & 0.22 & 0.18 \\
$\mbox{[Ti I/H] }$ & $+0.24$ & $-0.30$ & $-0.05$ & $+0.03$ & $-0.05$ & $+0.07$ & $-0.04$ & $+0.02$ & 0.31 & 0.05 \\
$\mbox{[Ti II/H]}$ & $+0.03$ & $-0.07$ & $+0.12$ & $-0.14$ & $-0.10$ & $+0.13$ & $+0.03$ & $-0.04$ & 0.21 & 0.07 \\
$\mbox{[Cr I/H] }$ & $+0.23$ & $-0.30$ & $-0.06$ & $+0.03$ & $-0.10$ & $+0.11$ & $-0.04$ & $+0.02$ & 0.33 & 0.02 \\
$\mbox{[Mn I/H] }$ & $+0.24$ & $-0.32$ & $-0.08$ & $+0.06$ & $-0.07$ & $+0.06$ & $-0.02$ & $+0.01$ & 0.34 & 0.08 \\
$\mbox{[Fe I/H] }$ & $+0.22$ & $-0.28$ & $-0.05$ & $+0.03$ & $-0.09$ & $+0.12$ & $-0.04$ & $+0.02$ & 0.31 & \nodata \\
$\mbox{[Fe II/H]}$ & $-0.03$ & $+0.00$ & $+0.14$ & $-0.15$ & $-0.09$ & $+0.12$ & $+0.03$ & $-0.04$ & 0.20 & \nodata \\
$\mbox{[Co I/H] }$ & $+0.26$ & $-0.35$ & $-0.07$ & $+0.03$ & $-0.22$ & $+0.26$ & $-0.07$ & $+0.03$ & 0.45 & 0.16 \\
$\mbox{[Ni I/H] }$ & $+0.20$ & $-0.24$ & $-0.02$ & $+0.02$ & $-0.05$ & $+0.07$ & $-0.03$ & $+0.01$ & 0.25 & 0.07 \\
$\mbox{[Sr II/H]}$ & $+0.04$ & $-0.09$ & $+0.05$ & $-0.14$ & $-0.11$ & $+0.11$ & $+0.00$ & $-0.06$ & 0.21 & 0.13 \\
$\mbox{[Ba II/H]}$ & $+0.09$ & $-0.10$ & $+0.10$ & $-0.12$ & $-0.07$ & $+0.09$ & $+0.02$ & $-0.04$ & 0.19 & 0.13 \\
\hline
TriII-40 & $+175$ & $-175$ & $+0.42$ & $-0.42$ & $+0.34$ & $-0.34$ & $+0.21$ & $-0.21$ & $$ & $$ \\
\hline
$\mbox{[Na I/H] }$ & $+0.20$ & $-0.23$ & $-0.05$ & $+0.04$ & $-0.06$ & $+0.08$ & $-0.01$ & $+0.01$ & 0.25 & 0.03 \\
$\mbox{[Mg I/H] }$ & $+0.22$ & $-0.22$ & $-0.11$ & $+0.11$ & $-0.11$ & $+0.10$ & $-0.01$ & $+0.01$ & 0.27 & 0.09 \\
$\mbox{[K I/H]  }$ & $+0.15$ & $-0.19$ & $-0.04$ & $+0.02$ & $-0.03$ & $+0.03$ & $-0.01$ & $+0.00$ & 0.20 & 0.09 \\
$\mbox{[Ca I/H] }$ & $+0.13$ & $-0.16$ & $-0.04$ & $+0.03$ & $-0.02$ & $+0.03$ & $-0.01$ & $+0.00$ & 0.17 & 0.11 \\
$\mbox{[Ti I/H] }$ & $+0.21$ & $-0.26$ & $-0.06$ & $+0.04$ & $-0.03$ & $+0.05$ & $-0.01$ & $+0.01$ & 0.27 & 0.04 \\
$\mbox{[Ti II/H]}$ & $+0.10$ & $-0.07$ & $+0.13$ & $-0.11$ & $-0.07$ & $+0.10$ & $+0.02$ & $-0.01$ & 0.19 & 0.08 \\
$\mbox{[Cr I/H] }$ & $+0.22$ & $-0.26$ & $-0.06$ & $+0.04$ & $-0.06$ & $+0.08$ & $-0.01$ & $+0.01$ & 0.28 & 0.02 \\
$\mbox{[Fe I/H] }$ & $+0.22$ & $-0.26$ & $-0.05$ & $+0.04$ & $-0.07$ & $+0.08$ & $-0.02$ & $+0.01$ & 0.28 & \nodata \\
$\mbox{[Fe II/H]}$ & $+0.03$ & $-0.01$ & $+0.14$ & $-0.13$ & $-0.06$ & $+0.08$ & $+0.01$ & $-0.01$ & 0.16 & \nodata \\
$\mbox{[Ni I/H] }$ & $+0.21$ & $-0.25$ & $-0.04$ & $+0.04$ & $-0.04$ & $+0.05$ & $-0.01$ & $+0.01$ & 0.26 & 0.03 \\
\hline
TriII-46 & $+200$ & $-200$ & $+0.50$ & $-0.50$ & $+0.50$ & $-0.50$ & $+0.30$ & $-0.30$ & $$ & $$ \\
\hline
$\mbox{[Na I/H] }$ & $+0.18$ & $-0.28$ & $-0.13$ & $+0.08$ & $-0.20$ & $+0.18$ & $-0.02$ & $-0.06$ & 0.37 & 0.09 \\
$\mbox{[Mg I/H] }$ & $+0.22$ & $-0.29$ & $-0.24$ & $+0.22$ & $-0.12$ & $+0.08$ & $+0.02$ & $-0.05$ & 0.40 & 0.23 \\
$\mbox{[Ca I/H] }$ & $+0.14$ & $-0.19$ & $-0.05$ & $+0.03$ & $-0.10$ & $+0.13$ & $-0.00$ & $-0.03$ & 0.24 & 0.11 \\
$\mbox{[Ti II/H]}$ & $+0.06$ & $-0.07$ & $+0.17$ & $-0.17$ & $-0.09$ & $+0.13$ & $+0.03$ & $-0.04$ & 0.23 & 0.24 \\
$\mbox{[Fe I/H] }$ & $+0.21$ & $-0.26$ & $-0.06$ & $+0.04$ & $-0.16$ & $+0.21$ & $-0.01$ & $-0.03$ & 0.34 & \nodata \\
$\mbox{[Fe II/H]}$ & $+0.04$ & $-0.10$ & $+0.03$ & $-0.05$ & $-0.28$ & $+0.21$ & $+0.03$ & $-0.06$ & 0.31 & \nodata \\
\enddata
\tablecomments{$\sigma_{\mbox{[X/H]}}$ is the quadrature sum of the maximum error for $\Delta\Teff$, $\Delta\logg$, $\Delta\nu_t$, and $\Delta\mbox{[Fe/H]}$.
$\sigma_{\mbox{[X/Fe]}}$ is the same sum but including the change in [Fe\,I/H] or [Fe\,II/H], depending on whether species X is neutral or ionized.
Correlations between stellar parameters were not considered.
Statistical uncertainties for both quantities are not included in this table, but are in Table~\ref{tbl:abunds}.}
\end{deluxetable*}

\end{document}